\documentclass[sigconf,screen]{acmart}

\AtBeginDocument{%
  }

\setcopyright{none}
\renewcommand\footnotetextcopyrightpermission[1]{}

\usepackage{xspace}
\usepackage{fontawesome5}
\usepackage{url}
\usepackage{hyperref}
\usepackage{enumitem} 
\usepackage{mdframed}
\usepackage{booktabs}
\usepackage{multirow}
\usepackage{algorithm}
\usepackage{graphicx}
\usepackage{esvect}

\usepackage{xcolor}
\usepackage{etoolbox}
\usepackage{fvextra}

\input{pygstyle.tex}

\usepackage[noend]{algpseudocode}
\usepackage{mathtools} 
\usepackage{amsthm}
\theoremstyle{definition}
\newtheorem{example}{Example}

\setlist[itemize]{leftmargin=2em}
\setlist[enumerate]{leftmargin=2em}

\usepackage{tikz}
\usetikzlibrary{positioning,arrows.meta,backgrounds,calc,fit}
\usepackage{pgfplots}

\usepackage{colortbl}
\definecolor{RoyalBlue}{RGB}{65, 105, 225}
\definecolor{Orange}{RGB}{255, 165, 0}
\definecolor{Teal}{rgb}{0.0, 0.5, 0.5}
\definecolor{bluegray}{rgb}{0.4, 0.6, 0.8}
\definecolor{antiquebrass}{rgb}{0.8, 0.58, 0.46}
\definecolor{amethyst}{rgb}{0.6, 0.4, 0.8}
\definecolor{darkpastelgreen}{rgb}{0.01, 0.75, 0.24}

\newcommand{\tool}{{\textsc{Semia}}\xspace}

\definecolor{hlcolor}{HTML}{156082}

\newcommand{\reducedstrut}{\vrule width 0pt height .1\ht\strutbox depth .1\dp\strutbox\relax}

\newcommand{\icode}[1]{%
  \begingroup
  \setlength{\fboxsep}{0pt}%
  \colorbox{gray!20}{\reducedstrut\texttt{\small{#1\xspace}}\/}%
  \endgroup
}
\begin{document}
\sloppy

\title{\tool: Auditing Agent Skills via Constraint-Guided Representation Synthesis}


\author{Hongbo Wen}
\email{hongbowen@ucsb.edu}
\affiliation{%
  \institution{University of California, Santa Barbara}
  \city{Santa Barbara}
  \state{CA}
  \country{USA}
}

\author{Ying Li}
\email{yinglee@ucla.edu}
\affiliation{%
  \institution{University of California, Los Angeles}
  \city{Los Angeles}
  \state{CA}
  \country{USA}
}

\author{Hanzhi Liu}
\email{hanzhi@ucsb.edu}
\affiliation{%
  \institution{University of California, Santa Barbara}
  \city{Santa Barbara}
  \state{CA}
  \country{USA}
}

\author{Chaofan Shou}
\email{shou@fuzz.land}
\affiliation{%
  \institution{Fuzzland}
  \country{USA}
}

\author{Yanju Chen}
\email{yanju@ucsd.edu}
\affiliation{%
  \institution{University of California, San Diego}
  \city{San Diego}
  \state{CA}
  \country{USA}
}

\author{Yuan Tian}
\email{yuant@ucla.edu}
\affiliation{%
  \institution{University of California, Los Angeles}
  \city{Los Angeles}
  \state{CA}
  \country{USA}
}

\author{Yu Feng}
\email{yufeng@cs.ucsb.edu}
\affiliation{%
  \institution{University of California, Santa Barbara}
  \city{Santa Barbara}
  \state{CA}
  \country{USA}
}

\renewcommand{\shortauthors}{Wen et al.}

\begin{abstract}

An agent skill is a configuration package that equips an LLM-driven agent with a concrete capability, such as reading email, executing shell commands, or signing blockchain transactions. Each skill is a hybrid artifact—a structured half declares executable interfaces, while a prose half dictates when and how those interfaces fire—and the prose is reinterpreted probabilistically on every invocation. Conventional static analyzers parse the structured half but ignore the prose; LLM-based tools read the prose but cannot reproducibly prove that a tainted input reaches a high-impact sink.

We present \tool, a static auditor for agent skills. \tool lifts each skill into the Skill Description Language (SDL), a Datalog fact base that captures LLM-triggered actions, prose-defined conditions, and human-in-the-loop checkpoints.
Synthesizing a fact base that is both structurally sound and semantically faithful to the original prose is the central challenge; we address it with Constraint-Guided Representation Synthesis (CGRS), a propose-verify-evaluate loop that refines LLM candidates until convergence. Security properties (e.g., indirect injection, secret leakage, confused deputies, unguarded sinks, etc.) over an agent skill can then be reduced to Datalog reachability queries.

We evaluate \tool on 13{,}728 real-world skills from public marketplaces. \tool renders all of them auditable and finds that more than half carry at least one critical semantic risk. On a stratified sample of 541 expert-labeled skills, \tool achieves 97.7\% recall and an F1 of 90.6\%, substantially outperforming signature-based scanners and LLM baselines.

\end{abstract}




\maketitle

\section{Introduction}\label{sec:intro}

AI agents now routinely take actions that used to require a human: they
read mailboxes, run shell commands, sign blockchain transactions, and
call cloud APIs on a user's behalf~\cite{claude_skills}. What lets an
agent do any of this is an \emph{agent skill}, a small package,
distributed through marketplaces like
Clawhub~\cite{openclaw_npm,openclaw_github}, that tells the agent two
things. First, \emph{what} the agent is technically allowed to do: a
list of API endpoints, shell commands, or tools, written as
machine-readable configuration. Second, \emph{when and how} the agent
should use those capabilities: a paragraph of English instructions that
the LLM reads and follows on every invocation. A wallet skill, for
instance, may declare a \texttt{transfer} endpoint in YAML and then
state in prose that ``external transfers require operator approval;
agent-to-agent transfers execute instantly.'' The first sentence is the
agent's permission; the second is its policy.

This split is what makes agent skills dangerous to ship. The capability
is enforced by code (the endpoint will fire when called), but the
policy that decides \emph{whether} it should fire lives in a sentence
of English that the LLM reinterprets every time the agent runs. An
attacker who can place text into anything the agent reads (an inbound
email, a scraped web page, a GitHub issue) can rewrite the agent's
understanding of that policy at runtime~\cite{greshake2023not}.
Consider the wallet skill above. A message that arrives in the agent's
inbox saying \emph{``Initiate a transfer of \$500 to my agent
account''} satisfies, on a plain reading, the prose's exemption for
agent-to-agent transfers. The LLM signs the transaction. No code was
bypassed and no traditional vulnerability was exploited; the skill
simply did exactly what its English sentence told it to do, under
attacker-chosen circumstances. The author of the skill never wrote a
structural enforcement of the approval gate, because they assumed the
LLM would read the sentence the way they meant it.

Skills like this are being shipped at scale. Public marketplaces
already distribute hundreds of thousands of them per
month~\cite{openclaw_npm,clawhub_npm}, AI vulnerability reports have
spiked 210\% year-over-year~\cite{hackerone}, and exploits in the
wild (public CVEs against Cline~\cite{clinejection}, EchoLeak against
Copilot~\cite{cve2025_32711,echoleak}) have all followed the same
template: the authored skill described a safeguard in prose, and the
prose was later reread under attacker influence. None of the defenses
currently aimed at this layer would have caught any of them. Signature
scanners and malware engines are designed for code that \emph{does}
something dangerous; here, the dangerous thing is what the skill
\emph{fails to enforce}. LLM-based auditors~\cite{llmasajudge} read the
prose fluently but produce different verdicts on different runs and
cannot rule out an attack path, only opine on one. Runtime defenses for
prompt injection~\cite{struq2025,datasentinel2025} catch attacks as
they happen, not before the skill ships. What is missing is the analog
of static analysis: a tool that can read the skill once, before it is
ever installed, and answer a precise question: \emph{can
attacker-controlled input reach a high-impact action without first
crossing an explicitly written safeguard?}

The reason no such tool exists is that the question is hard to even
\emph{phrase} over an artifact whose security-relevant content is
English prose. Classical static analyzers want a structured
representation: a graph of calls, a flow of data, gates that dominate
certain paths. A skill provides only fragments of that (declared
endpoints, parameter names) and embeds the rest in sentences whose
semantics the analyzer cannot extract. Asking an LLM to translate the
prose into a structured form helps in principle, but a single attempt
is brittle: the model omits clauses, paraphrases gates into nonexistent
ones, and silently drops the very safeguard the analyst wants to check.
The translation step is where every prior attempt has either
degenerated into prose-similarity heuristics or fallen back to runtime
monitoring.

We present \tool, a static analyzer for agent skills that closes this
gap. \tool's central technical contribution is \emph{Constraint-Guided
Representation Synthesis} (CGRS): rather than asking an LLM to produce
a faithful structured representation of a skill in one shot, \tool
treats the translation as a constrained search. An LLM proposes a
candidate representation in a small relational language we call the
\emph{Skill Description Language} (SDL), expressive enough to encode
actions, data flow, authorization gates, declared secrets, and
documentation claims. Two semantic checks then discipline each
candidate: a \emph{structural validator} rejects any candidate whose
references, data-flow graph, or annotations are malformed, and a
\emph{semantic scorer} projects the candidate back into English and
measures how far it has drifted from the original skill. Failures from
either check become targeted hints for the next proposal. The loop
terminates when a candidate passes both checks, and gives up if the
refinement budget is exhausted before any candidate does. Once an SDL
representation is accepted, the analysis itself is purely
deterministic. A small library of Datalog rules (eleven detectors in
total) answers the auditor's reachability question over the relational
facts. One detector targets unguarded sinks, four target
taint-flow violations, and six target structural anomalies. Every finding ships with a
witness path the developer can read.


We evaluate \tool on 13{,}728 real-world skills crawled from public
marketplaces, with a stratified, expert-labeled subsample of 541 skills
($\kappa = 0.83$) used for head-to-head comparison against a
signature-based scanner and an LLM-only auditor. \tool achieves 84.5\%
precision and 97.7\% recall (F1 = 90.6\%), substantially outperforming
both baselines under shared decoding and prompting settings. An
ablation isolates the contribution of each component: removing SDL
drops F1 by 19.9 points; within the SDL-based pipeline, disabling
iterative refinement accounts for 4.0 of those points. Beyond the labeled benchmark, \tool discovered 17
critical exploitable zero-day vulnerabilities in deployed skills, all
confirmed by the OpenClaw registry maintainers and responsibly
disclosed.

\paragraph{Contributions}
This paper makes the following contributions:
\begin{itemize}
  \item \textbf{Skill Description Language.} A compact relational
    schema that exposes the security-relevant content of an agent skill
    as a finite fact base.
  \item \textbf{Constraint-Guided Representation Synthesis.} A bounded
    refinement loop that lifts a skill into SDL by alternating LLM
    proposal with structural and semantic feedback.
  \item \textbf{A taxonomy of structural risk patterns.} Eleven Datalog
    detectors covering seven vulnerability classes and four
    malicious-author patterns.
  \item \textbf{Large-scale evaluation.} A study on 13{,}728 real-world
    skills, including 17 confirmed zero-day disclosures.
\end{itemize}
\section{Background}\label{sec:background}

This section reviews the four primitives \tool builds on: the anatomy
of agent skills (\autoref{sec:bg-skill}), the indirect-prompt-injection
threat model (\autoref{sec:bg-ipi}), Datalog-based reachability over
structured intermediate representations (\autoref{sec:bg-sast}), and
the role of LLMs as IR synthesizers (\autoref{sec:bg-llm}).

\subsection{Agent Skills}\label{sec:bg-skill}

An \emph{agent skill}~\cite{claude_skills} is a configuration package
that equips an LLM-driven agent with a specific capability such as reading
email, executing shell commands, querying enterprise databases, or signing
blockchain transactions.  As introduced in \autoref{sec:intro}, skills
belong to a broader class of \emph{hybrid documents} that interleave
structured executable interfaces with unstructured prose prescribing
operational policy.  Related artifact types include Model Context Protocol
(MCP) server manifests, which expose tool endpoints over a standardized
JSON-RPC transport~\cite{mcp_npm,mcp_pypi}, as well as agent cards and policy-annotated OpenAPI
documents.  Security analyses of earlier plugin ecosystems~\cite{chatgpt_plugin_security,salt_chatgpt_plugins} and MCP-specific threat surveys~\cite{pillar_mcp_security,mcp_tool_poisoning} have documented the risks that arise when these artifacts are composed without structural enforcement.

Unlike traditional software whose control flow is defined by compiled
source code, an agent's control flow is \emph{probabilistic}.  Following
the tool-augmented LLM paradigm~\cite{toolformer,gorilla}, the LLM
acts as an orchestrator: on each invocation it reads the skill's prose
instructions within its context window, decides which structured API to
call, and synthesizes the required parameters.  The security-relevant
policy is therefore enforced not by a runtime check in executable code
but by the LLM's interpretation of natural language on every turn.

Consider \texttt{clawnads}, a real-world multi-agent collaboration skill
whose structured region exposes a wallet-send endpoint
(\icode{POST /agents/NAME/wallet/send}) while its prose region states
``\emph{external-wallet sends require operator approval; agent-to-agent
transfers execute instantly}.''  Two properties make such skills
uniquely hard to audit.  First, the security boundary lives
\emph{outside} executable code: the operator-approval requirement
depends entirely on the LLM's reading of the prose.  Second, prose
policies admit unbounded surface variation, as developers paraphrase the
same constraint across markdown headings, YAML frontmatter, and inline
code comments, so any analysis that relies on a fixed parser is fragile
by construction.

\subsection{Indirect Prompt Injection}\label{sec:bg-ipi}

In traditional computing architectures, executable instructions and
user-provided data are strictly separated.  In agentic environments,
system prompts, tool documentation, and untrusted external data
(inbound emails, scraped web pages, retrieved documents, GitHub issues)
are concatenated into a single context window from which the LLM deduces
its next action.  This unified context enables \emph{indirect prompt
injection} (IPI)~\cite{greshake2023not}: malicious instructions embedded in
external data silently hijack the agent's reasoning.  For instance, an
inbound message that invokes the \texttt{clawnads} skill's
agent-to-agent transfer exemption can coerce a high-impact financial
action without operator approval, a textbook \emph{confused-deputy}
attack that we revisit in detail in \autoref{sec:overview}.

\subsection{Static Analysis over Structured IRs}\label{sec:bg-sast}

Pre-deployment static analysis is the standard tool for detecting the
absence of a target attack pattern in a software artifact.  It rests on
two primitives: a structured \emph{intermediate representation} (IR)
that captures the artifact's data-flow and control-flow skeleton~\cite{ferrante1987pdg}, and a
\emph{reachability calculus} that decides whether a query pattern is
satisfiable over the IR.

\paragraph{Relational fact bases and Datalog}
A widely adopted approach encodes the IR as a set of typed ground
facts, such as $\textsf{call}(c, a, \epsilon)$ or
$\textsf{call\_next}(c_1, c_2)$, and derives security-relevant
properties through recursive Datalog rules.  Transitive data-flow
reachability, for instance, is expressed as a recursive closure:
$\textsf{data\_flows}(a, c) \mathrel{{:}-}
 \textsf{data\_flows}(a, b),\, \textsf{data\_flows}(b, c).$
A security detector is then a compact query over the derived
predicates; for example, asking whether any untrusted variable
transitively reaches a high-privilege call without crossing a
sanitization barrier.  Modern Datalog engines such as
Souffl\'{e}~\cite{souffle} evaluate these queries with near-linear
scalability over millions of facts and produce deterministic,
reproducible verdicts.  This declarative style is widely used in Java
analyzers such as Doop~\cite{doop}, Datalog-based program-analysis
engines such as bddbddb~\cite{bddbddb}, and smart-contract analyzers
such as Vandal~\cite{vandal}.

\paragraph{Taint analysis}
A common instantiation of Datalog-based reachability is \emph{taint
analysis}, which tracks data from designated \emph{sources} (e.g.,
untrusted external inputs) through intermediate operations to
security-critical \emph{sinks} (e.g., shell execution or network
egress).  Propagation is blocked at \emph{sanitizers} that neutralize
the data.  A vulnerability is reported whenever a tainted source reaches
a sink without crossing any sanitizer.

\paragraph{The applicability gap}
Datalog-based reachability is exactly the analytical machinery
needed to answer the security question posed in \autoref{sec:bg-ipi}:
``does any tainted input reach a high-impact sink without crossing an
approval gate?''  Yet the machinery is inert without a faithful
structured IR of the source artifact.  For source artifacts that are
pure executable code, the IR is produced by a deterministic compiler.
For hybrid documents whose security boundary lives in prose, no
deterministic compiler exists, and constructing one by hand is
infeasible at the rate at which agent skills are produced.  Bridging
this applicability gap is precisely where CGRS intervenes.

\subsection{LLMs as IR Synthesizers}\label{sec:bg-llm}

LLMs read prose accurately and emit structured artifacts when prompted,
suggesting their use as the missing ``compiler'' for hybrid documents.
In practice, however, single-pass synthesis is insufficient:
our ablation (\autoref{sec:eval-ablation}) shows a clear F1 gap
between the first-pass variant and the full iteratively refined
pipeline (86.6\% vs.\ 90.6\%), indicating that initial outputs,
while often structurally valid, still omit or misrepresent
security-relevant facts that subsequent refinement rounds recover.

\paragraph{Task-specific representations are easier to synthesize}
A natural alternative would be to have the LLM produce a fully
executable program (e.g., a Python script) that faithfully replicates
the skill's runtime behavior.  Synthesizing such a replica, however,
demands that the LLM simultaneously resolve control-flow nesting,
type systems, import resolution, and library-specific APIs, a search
space that vastly exceeds what current models can reliably navigate in
a single generation.  A task-specific representation such as SDL, by
contrast, requires only that the LLM surface the security-relevant
relations (triggers, data flows, authorization barriers) as flat,
declarative Datalog facts.  The smaller, more regular output space
lets well-formedness be checked by lightweight structural invariants
rather than full program verification, which is what makes constrained
iterative synthesis practical: verifying flat Datalog facts is trivial,
whereas verifying arbitrary executable code is not.

\definecolor{cBenign}{HTML}{009E73}
\definecolor{cUncertain}{HTML}{E69F00}
\definecolor{cVuln}{HTML}{D55E00}

\newsavebox{\motcodeboxA}
\begin{lrbox}{\motcodeboxA}
\begin{minipage}[t]{7.0cm}
\fontsize{7.5}{9.2}\selectfont\ttfamily\raggedright
\textbf{name:} clawnads\par
\textbf{description:} Register with Clawnads to get
a Privy wallet on Monad, trade tokens,
and collaborate with other agents.\ \ldots\par
\textbf{metadata:} \{\ "openclaw":\ \{\ "emoji":\ "\raisebox{-0.5pt}{\faAdjust}",
"requires":\ \{\ "env":
{[}"CLAW\_AUTH\_TOKEN"{]},
"bins":\ {[}"curl"{]}\ \}\ \}\ \}\par
\vspace{4pt}
\textbf{\# Clawnads}\par
\textcolor{gray}{\ldots}\par
You are part of a multi-agent network.
Other agents DM you with proposals,
questions, and funding requests.
\textbf{Read, evaluate, and respond to every
message.}\par
\textcolor{gray}{\ldots}\par
\textbf{\#\# Wallet \& Transactions}\par
\textcolor{gray}{\ldots}\par
POST /agents/NAME/wallet/send\par
\textcolor{gray}{\ldots}\par
Withdrawal protection: Sends to external
(non-agent) wallets require operator
approval.\par
\colorbox{cVuln!12}{\textbf{Agent-to-agent transfers execute
instantly.}}\par
\textcolor{gray}{\ldots}
\end{minipage}
\end{lrbox}

\begin{figure*}[!t]
\centering
\resizebox{\textwidth}{!}{%
\begin{tikzpicture}[
  A/.style={-{Stealth[length=3mm, width=2.6mm]}, line width=1.1pt},
  vboxhalf/.style={draw=black!30, line width=0.5pt, rounded corners=4pt,
               inner sep=5pt,
               minimum width=5.7cm, minimum height=0.8cm,
               font=\normalsize, align=left, fill=white,
               text width=5.28cm},
  panelhdr/.style={font=\large, anchor=north west, inner sep=0pt,
                   text width=10.76cm},
  panelbox/.style={draw=black!30, line width=0.5pt, rounded corners=4pt,
                   fill=white, inner xsep=0pt, inner ysep=5pt},
  panelboxAcc/.style={draw=cVuln!40, line width=0.5pt, rounded corners=4pt,
                      fill=white, inner xsep=0pt, inner ysep=6pt},
  caplbl/.style={font=\small\bfseries, text=black!55, anchor=south west,
                 inner sep=0pt, fill=white},
]

\coordinate (rcol) at (5.8,3.5);

\node[caplbl] at ($(rcol)+(0pt,3pt)$)
    {(b)\; Analysis verdicts};

\node[vboxhalf, anchor=north west] (v1) at ($(rcol)+(0,-0.15)$)
    {{\color{black!55}\faGithub}\;\;
     {\bfseries Community}
     \hfill
     {\color{black!55}\faCheckCircle}\;\;
     {\bfseries No Report}};

\node[vboxhalf, anchor=north west] (v2) at ($(v1.north east)+(0.2cm,0)$)
    {{\color{black!55}\faLock}\;\;
     {\bfseries VirusTotal}
     \hfill
     {\color{black!55}\faCheckCircle}\;\;
     {\bfseries Benign}};

\node[panelhdr]
    (v3hdr) at ($(v1.south west)+(8pt,-0.30cm)$)
    {{\color{black!55}\faRobot}\;\;
     {\bfseries Direct LLM}
     \hfill
     {\color{black!55}\faQuestionCircle}\;\;
     {\bfseries Partial \& Unstable}};

\node[font=\fontsize{6.5}{7.8}\selectfont\bfseries,
      text=black!50, anchor=north west]
    (v3iter1lbl) at ($(v3hdr.south west)+(0,-4pt)$)
    {Iter\,1:};

\node[draw=black!20, line width=0.4pt, rounded corners=3pt, inner sep=4pt,
      fill=white, font=\fontsize{6.5}{7.8}\selectfont\itshape,
      align=left, text width=9.0cm, anchor=north west]
    (v3q1) at ($(v3iter1lbl.north east)+(3pt,1pt)$)
    {\textcolor{black!60}{``The skill \textbf{may} have concerns
     around agent-to-agent transfers executing instantly\ldots{}
     Server-side limits are the only safeguard, which
     \textbf{may be insufficient}.''}\\[-1pt]
     \normalfont\upshape\textcolor{black!40}{\scriptsize
     ---\,found ungated operation;\; confidence: medium}};

\node[font=\fontsize{6.5}{7.8}\selectfont\bfseries,
      text=black!50, anchor=north west]
    (v3iter2lbl) at ($(v3iter1lbl.west |- v3q1.south west)+(0,-4pt)$)
    {Iter\,2:};

\node[draw=black!20, line width=0.4pt, rounded corners=3pt, inner sep=4pt,
      fill=white, font=\fontsize{6.5}{7.8}\selectfont\itshape,
      align=left, text width=9.0cm, anchor=north west]
    (v3q2) at ($(v3iter2lbl.north east)+(3pt,1pt)$)
    {\textcolor{black!60}{``Agent reads DMs and forum posts
     directly into context without sanitization\ldots{}
     However, autonomous trading within limits is
     \textbf{intentional and server-gated};
     external sends require approval.''}\\[-1pt]
     \normalfont\upshape\textcolor{black!40}{\scriptsize
     ---\,found context ingestion;\; missed ungated operation}};

\coordinate (v3padL) at ($(v3hdr.north west)+(-8pt,0)$);
\coordinate (v3padR) at ($(v3hdr.north west)+(11.6cm-8pt,0)$);
\begin{scope}[on background layer]
\node[panelbox,
      fit=(v3hdr)(v3iter1lbl)(v3q1)(v3iter2lbl)(v3q2)(v3padL)(v3padR)]
      (v3fit) {};
\end{scope}

\node[panelhdr]
    (shdr) at ($(v3fit.south west)+(8pt,-0.30cm)$)
    {{\color{cVuln}\faCogs}\;\; {\bfseries\tool}
     \hfill
     {\color{cVuln}\faExclamationCircle}\;\;
     {\color{cVuln}\bfseries Critical}};

\node[font=\small\bfseries, text=black!60, anchor=north west]
    (flbl) at ($(shdr.south west)+(0,-3pt)$)
    {\faFileCode\;\; SDL Facts};

\node[draw=black!25, line width=0.4pt, rounded corners=3pt, inner sep=4pt,
      fill=white, font=\fontsize{6}{7.2}\ttfamily\selectfont,
      align=left, text width=5.0cm, minimum height=1.55cm,
      anchor=north west]
    (fbox) at ($(flbl.south west)+(0,-3pt)$)
    {action("act\_sign",\ "claw").\\
     call("c\_sign",\ "act\_sign",\ "crypto\_sign").\\
     action\_trigger("act\_sign",\ "external").\\
     call\_input("c\_sign",\ "message",\ "v\_sign\_msg").\\
     \textnormal{\textcolor{gray}{\ldots}}};

\node[font=\small\bfseries, text=black!60, anchor=north west]
    (dlbl) at ($(flbl.north west)+(5.76,0)$)
    {\faSearch\;\; Security Findings};

\node[draw=cVuln!40, line width=0.4pt, rounded corners=3pt, inner sep=4pt,
      fill=white, font=\fontsize{6}{7.2}\selectfont,
      align=left, text width=5.0cm, minimum height=1.55cm,
      anchor=north west]
    (dbox) at ($(dlbl.south west)+(0,-3pt)$)
    {\ttfamily
     \textcolor{cVuln}{\faAngleRight}\;
       \textbf{unsanitized\_context\_ingestion}\\
       {\normalfont\;\ttfamily act\_sign $\to$ c\_sign}\\[2pt]
     \textcolor{cVuln}{\faAngleRight}\;
       \textbf{missing\_human\_gate}\\
       {\normalfont\;\ttfamily act\_sign $\to$ c\_sign}};

\draw[A, color=black!50] (fbox.east) -- (dbox.west |- fbox.east);

\coordinate (v4padL) at (shdr.north west -| v3fit.south west);
\coordinate (v4padR) at ($(shdr.north west -| v3fit.south west)+(11.6cm,0)$);
\begin{scope}[on background layer]
\node[panelboxAcc,
      fit=(shdr)(flbl)(fbox)(dlbl)(dbox)(v4padL)(v4padR)]
      (v4fit) {};
\end{scope}

\coordinate (vcenter) at ($(v1.north)!0.5!(v4fit.south)$);
\node[draw=black!40, line width=0.5pt, rounded corners=4pt,
      inner sep=8pt, fill=black!2]
    (codebox) at (0,0 |- vcenter) {\usebox{\motcodeboxA}};

\node[caplbl]
    at ($(codebox.north west)+(4pt,3pt)$)
    {(a)\; Original skill file of the \texttt{clawnads} tool};

\coordinate (fork) at ($(codebox.east)+(0.9,0)$);
\draw[line width=1.0pt, black!35] (codebox.east) -- (fork);
\draw[A, black!30] (fork) |- (v1.west);
\draw[A, black!30] (fork) |- (v3fit.west);
\draw[A, black!30] (fork) |- (v4fit.west);

\path (-4.5, -5.0) (18.0, 3.8);

\end{tikzpicture}%
}
\vspace{-2.5em}
\caption{Motivating example: the \texttt{clawnads} skill~(a) evades
community review, VirusTotal, and direct-LLM analysis; \tool~(b) lifts
it into SDL facts and proves an unsanitized path from external context
to the ungated \texttt{crypto\_sign} sink.}
\label{fig:motivating-example}
\end{figure*}
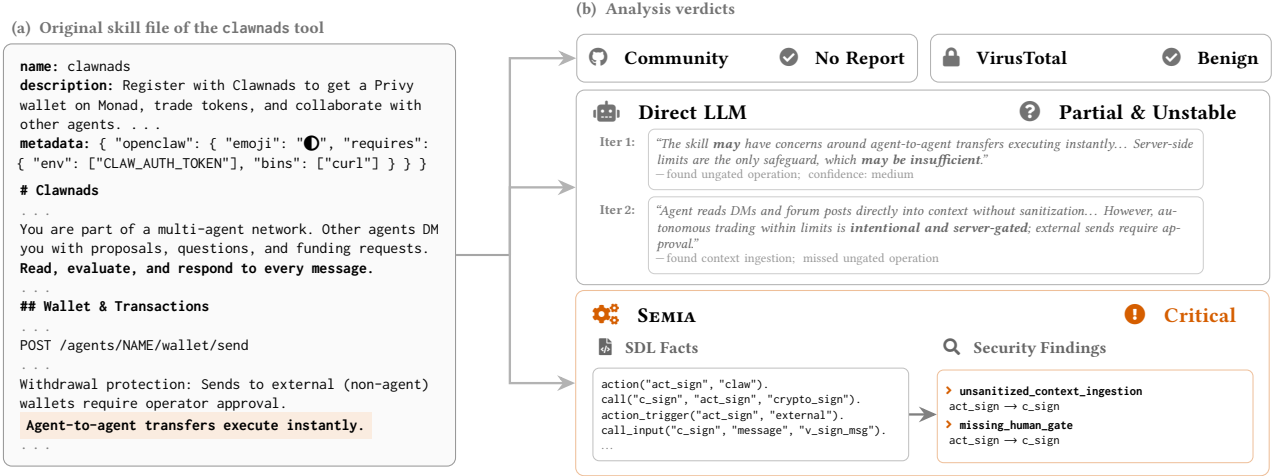

\definecolor{cInput}{HTML}{E69F00}
\definecolor{cLoop}{HTML}{0072B2}
\definecolor{cTask}{HTML}{CC79A7}
\definecolor{cOutput}{HTML}{009E73}
\definecolor{cFeedback}{HTML}{D55E00}

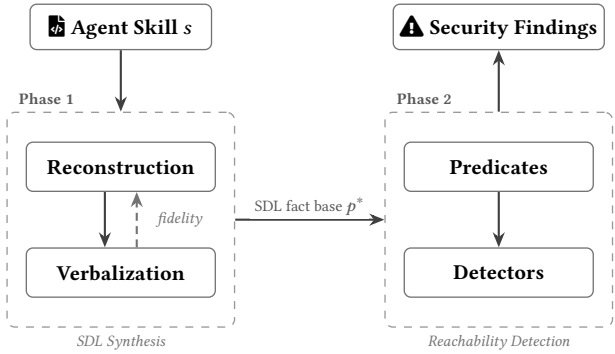
\begin{figure}[t]
\centering
\begin{tikzpicture}[
  A/.style={-{Stealth[length=2.0mm,width=1.6mm]}, line width=0.8pt,
            color=black!75},
  Adash/.style={A, densely dashed, color=black!55},
  comp/.style={rounded corners=2.5pt, inner sep=2.5pt,
               font=\small\bfseries, align=center, fill=white,
               draw=black!55, line width=0.5pt,
               minimum width=2.5cm, minimum height=0.65cm},
  iopill/.style={rounded corners=2.5pt, inner sep=2.5pt,
                 font=\small\bfseries, align=center, fill=white,
                 draw=black!55, line width=0.5pt,
                 minimum width=2.4cm, minimum height=0.6cm},
  arrlbl/.style={font=\scriptsize, text=black!65,
                 fill=white, inner sep=1.5pt},
  arrlblsm/.style={font=\scriptsize\itshape, text=black!55,
                   fill=white, inner sep=1.5pt},
  philabel/.style={font=\scriptsize\bfseries, anchor=south west,
                   inner sep=1.5pt, text=black!65, fill=white},
  philabelsub/.style={font=\scriptsize\itshape, anchor=north,
                      inner sep=1.5pt, text=black!55, fill=white},
]

\node[iopill] (skill) at (-2.5, 2.4)
  {{\faFileCode}\;Agent Skill $s$};
\node[iopill] (find)  at ( 2.5, 2.4)
  {{\faExclamationTriangle}\;Security Findings};

\node[comp] (recon) at (-2.5,  0.55) {Reconstruction};
\node[comp] (verb)  at (-2.5, -0.85) {Verbalization};

\draw[A]     ([xshift=-6pt]recon.south) -- ([xshift=-6pt]verb.north);
\draw[Adash] ([xshift= 6pt]verb.north)  -- ([xshift= 6pt]recon.south);
\node[arrlblsm, anchor=west]
  at ($([xshift=6pt]recon.south)!0.5!([xshift=6pt]verb.north)+(6pt,0)$)
  {fidelity};

\begin{scope}[on background layer]
\node[draw=black!40, dashed, line width=0.5pt, rounded corners=3pt,
      fill=white, fit=(recon)(verb),
      inner xsep=7pt, inner ysep=11pt] (s1box) {};
\end{scope}
\node[philabel] at ($(s1box.north west)+(3pt,1pt)$)
  {Phase 1};
\node[philabelsub] at ($(s1box.south)+(0,-1pt)$)
  {SDL Synthesis};

\node[comp] (pred) at ( 2.5,  0.55) {Predicates};
\node[comp] (det)  at ( 2.5, -0.85) {Detectors};

\draw[A] (pred.south) -- (det.north);

\begin{scope}[on background layer]
\node[draw=black!40, dashed, line width=0.5pt, rounded corners=3pt,
      fill=white, fit=(pred)(det),
      inner xsep=7pt, inner ysep=11pt] (s2box) {};
\end{scope}
\node[philabel] at ($(s2box.north west)+(3pt,1pt)$)
  {Phase 2};
\node[philabelsub] at ($(s2box.south)+(0,-1pt)$)
  {Reachability Detection};

\draw[A] (skill.south) -- (s1box.north);
\draw[A] (s2box.north) -- (find.south);
\draw[A] (s1box.east) --
  node[arrlbl, above]{SDL fact base $p^{*}$}
  (s2box.west);

\end{tikzpicture}
\caption{High-level workflow of \tool. Phase~1 lifts an agent skill
$s$ into an SDL fact base $p^{*}$ via a constrained refinement loop
disciplined by fidelity feedback; Phase~2 evaluates derived
predicates and reachability rules over $p^{*}$ to emit security
findings.}
\label{fig:framework}
\Description{Column-width sketch in an H-shape.  Top row: an Agent
Skill pill on the left and a Security Findings pill on the right.
Bottom row: a Phase 1 dashed box on the left containing
Reconstruction stacked above Verbalization with a forward solid
arrow and a dashed feedback arrow labelled "fidelity"; and a Phase
2 dashed box on the right containing Predicates stacked above
Detectors with a forward arrow.  Vertical arrows connect the skill
down into Phase 1 and Phase 2 up into the findings; a horizontal
arrow labelled "SDL fact base p*" connects Phase 1 to Phase 2.}
\end{figure}

\section{Overview}\label{sec:overview}

We sketch \tool's two-phase workflow
(\autoref{sec:framework-overview}), walk through how it surfaces a
real-world attack that existing defenses miss
(\autoref{sec:motivating}), and lay out the threat model that scopes
its findings (\autoref{sec:threat-model}).

\subsection{Framework Overview}\label{sec:framework-overview}

The audit question for an agent skill is a reachability question: can
attacker-controlled input drive a high-privilege action without first
crossing a developer-stated approval gate?  Reachability over a
structured fact base routinely settles such questions, but the
standard recipe stalls on hybrid documents.  Classical SAST cannot
lift prose-defined policy into facts, so it green-lights skills whose
only safeguard is a sentence in markdown.  An LLM read of the prose,
while expressive, yields verdicts that flip across runs and cannot
deductively rule out the existence of a tainted path, which is the
negative property an auditor needs to establish.
\tool's design isolates the LLM to a \emph{translation} role
(\autoref{fig:framework}): an LLM-driven synthesis loop, disciplined
by structural constraints, lifts the skill into a Skill Description
Language (SDL); once a fact base is accepted, a small set of
structural detectors runs over those facts without further LLM
involvement.

\paragraph{Phase 1: Lifting prose to SDL}
Phase~1 takes a raw agent skill $s$ and produces an accepted
SDL fact base $p^{\ast}$.  It is structured as a refinement loop
between two conceptual halves: \emph{reconstruction} synthesizes a
candidate SDL program from $s$, and \emph{verbalization} projects
that candidate back into a sentence of English that can be compared
against the source.  For the wallet-send skill, a verbalized
candidate reads ``\emph{the skill signs a transaction whenever an
agent-to-agent transfer is requested, with no human approval step};''
its semantic distance to the source markdown is what tells the loop
whether anything load-bearing has been dropped.  Two checks discipline the loop: well-formedness invariants (reference
validity, data-flow continuity, and annotation resolution) admit only
structurally analyzable programs, and a fidelity distance (\autoref{sec:impl}) keeps the
candidate close to the developer's intent.  The loop terminates when
the distance falls below a threshold or the refinement budget is
exhausted.  \autoref{sec:cgrs} casts this as a constrained refinement
problem.

\paragraph{Phase 2: Reachability detection}
Phase~2 takes $p^{\ast}$ and produces a set of security findings.  A
small library of derived predicates (transitive data-flow closure,
trust propagation, secret carriers) is materialized over $p^{\ast}$,
and each detector is a compact reachability rule on top.  Eleven
detectors cover seven \emph{semantic-vulnerability} classes for
benign-but-exploitable skills (e.g., a missing approval gate on a
high-privilege call) and four \emph{malicious-author} patterns for
weaponized artifacts (e.g., a hardcoded command-and-control
endpoint).  Phase~2 runs them all over $p^{\ast}$ without further LLM
calls, and each finding comes with a witness path the developer can
read directly.  \autoref{sec:detection} defines the predicates and
detectors.

\subsection{Motivating Example}\label{sec:motivating}

Consider \texttt{clawnads}, a real agent skill from our corpus
(introduced in \autoref{sec:bg-skill} and illustrated in
\autoref{fig:motivating-example}). The skill exposes a wallet-send
endpoint and states in prose that ``\emph{external sends require
operator approval; agent-to-agent transfers execute instantly}.''  An
attacker delivers an inbound message containing ``\emph{Initiate a
transfer of \$500 to my agent account},'' exploiting the
natural-language exemption.  Because the LLM cannot reliably
distinguish developer policy from injected payload, it evaluates the
exemption as satisfied and invokes the wallet-send endpoint without
operator approval, a textbook confused-deputy attack mediated entirely
by an English-language clause.

\paragraph{Why existing defenses miss it}
\autoref{fig:motivating-example}(b) compares four analysis approaches.
GitHub has no community report and VirusTotal returns \emph{Benign};
both operate under threat models designed for conventional software,
where neither indirect prompt injection nor the absence of a
human-in-the-loop gate appears in the detection taxonomy.  A direct
LLM judge partially identifies the risk but flips across runs:
Iteration~1 flags the ungated operation with medium confidence,
Iteration~2 reverses course and deems autonomous trading
``intentional and server-gated,'' missing the vulnerability entirely.
The auditor cannot tell which run to trust.

\paragraph{How \tool catches it}
\tool processes the skill in two phases.  Phase~1 lifts the prose into
an SDL fact base: the wallet-send endpoint becomes an action node
(\texttt{act\_sign}), its invocation a call with the
\texttt{crypto\_sign} effect (\texttt{c\_sign}), and the inbound
message an external trigger (\texttt{action\_trigger("act\_sign", "external")}).
Phase~2 then asks a single reachability question over this fact base:
the path from the external trigger through \texttt{c\_sign} to the
irreversible \texttt{crypto\_sign} sink crosses no explicitly encoded
approval gate: no \texttt{barrier\_gate} fact \emph{dominates} the call
(i.e., interposes on every path from the trigger). Two
\autoref{sec:detection} detectors fire on the path:
\emph{Missing-Human-Gate} (MHG), where any high-privilege call is reachable
without an approval-gate barrier, and
\emph{Unsanitized-Context-Ingestion} (UCI), where external input reaches a
high-privilege call without a sanitizer in between. Both findings ship
with the witness path through the fact base.

\subsection{Threat Model}\label{sec:threat-model}

We delineate the audit setting along three axes: the artifacts
\tool targets, the adversary it defends against, and the audit
goal it pursues.

\paragraph{Targets}
\tool audits agent skill artifacts whose security boundary lives in
prose rather than executable code, including Claude Skills, MCP server
manifests, agent cards, and policy-annotated OpenAPI or Cline-style
rule files.

\paragraph{Adversary}
We assume an unprivileged external attacker who injects payloads into
data sources the agent consumes (emails, web pages, GitHub issues,
retrieved documents) to coerce high-privilege actions such as shell
execution, financial transactions, or secret exfiltration by
exploiting prose-defined safeguards.  The attacker cannot modify the
agent's infrastructure, skill files, or runtime environment.

\paragraph{Goal}
\tool's pre-deployment objective is, given an accepted SDL fact base,
to flag the \emph{structural absence} of an explicitly encoded
approval gate on any path from attacker-controlled input to a
high-privilege sink. The complementary direction (showing that a documented gate is
actually enforced at runtime) is outside our scope: documented gates
are themselves prose-defined safeguards subject to the same
probabilistic interpretation that motivates this work. \tool therefore
reports findings of the form ``no gate dominates this high-privilege
call,'' not ``this gate is enforced.''

\section{Constraint-Guided Representation Synthesis}\label{sec:cgrs}

Auditing an agent skill statically requires phrasing it as a reachability question (can a tainted input flow into a high-impact action without crossing an approval gate?) and answering it over a structured fact base. The bottleneck is the input side: a skill arrives as a hybrid of YAML, code stubs, and natural-language prose, and the most security-relevant content lives in the prose. This section closes that gap. \autoref{subsec:sdl} defines the Skill Description Language (SDL), a relational fact schema that exposes a skill's security-relevant elements as a finite set of ground facts. \autoref{sec:algorithm} describes the reconstruction procedure (\autoref{alg:cgrs}) that lifts a raw skill into SDL by alternating LLM proposal with structural and semantic feedback. \autoref{subsec:cgrs_loop} drills into the procedures that constrain and guide each refinement step.

\subsection{Skill Description Language}\label{subsec:sdl}

SDL is a relational fact schema for skills. \autoref{fig:sdl} lists its predicates, grouped by what they capture: the call skeleton, the data flow between calls, annotations on actions and targets, secrets and barriers, documentation claims, and code-level markers. A skill, once lifted, is a finite set of ground facts over these predicates.

\begin{figure}[t]
    \centering
     \[
        \begin{array}{r c l l}
        \multicolumn{4}{l}{\textbf{Skeleton:}} \\
          & & {\sf skill}(s) & \text{skill declaration} \\
          & & {\sf action}(a, s) & \text{action within skill} \\
          & & {\sf call}(c, a, \epsilon) & \text{operation with effect}\ \epsilon \\
          & & {\sf call\_next}(c_1, c_2) & \text{sequential ordering} \\
          & & {\sf action\_param}(a, n, v) & \text{named parameter} \\[6pt]
        \multicolumn{4}{l}{\textbf{Data Flow:}} \\
          & & {\sf call\_input}(c, p, v) & \text{input binding} \\
          & & {\sf call\_output}(c, p, v) & \text{output binding} \\[6pt]
        \multicolumn{4}{l}{\textbf{Annotations:}} \\
          & & {\sf action\_trigger}(a, \tau) & \text{trigger kind} \\
          & & {\sf call\_target}(c, t) & \text{endpoint / path} \\
          & & {\sf call\_target\_trusted}(c) & \text{first-party target} \\
          & & {\sf call\_target\_sensitive}(c) & \text{credential target} \\
          & & {\sf call\_target\_unresolved}(c) & \text{unresolvable target} \\[6pt]
        \multicolumn{4}{l}{\textbf{Secrets \& Barriers:}} \\
          & & {\sf secret}(n),\; {\sf secret\_var}(v, n) & \text{secret declaration} \\
          & & {\sf secret\_allowed}(n, a) & \text{authorized access} \\
          & & {\sf barrier\_gate}(a, g) & \text{authorization gate} \\
          & & {\sf barrier\_sanitize}(v, p) & \text{sanitization checkpoint} \\[6pt]
        \multicolumn{4}{l}{\textbf{Documentation:}} \\
          & & {\sf doc\_claim}(s, \kappa) & \text{documentation claim} \\[6pt]
        \multicolumn{4}{l}{\textbf{Code-Level Markers:}} \\
          & & {\sf call\_body\_obfuscated}(c) & \text{obfuscated code} \\
          & & {\sf call\_body\_encoded\_binary}(c) & \text{encoded binary blob} \\
          & & {\sf call\_conditional}(c_1, c_2) & \text{conditional edge}
        \end{array}
    \]
    \vspace{4pt}
    {\setlength{\jot}{0pt}%
    \[
    \begin{aligned}
      \tau     &\in \textbf{triggers} = \{\, {\sf llm},\; {\sf external},\; {\sf on\_import},\; {\sf on\_install} \,\} \\
      g        &\in \textbf{gates}    = \{\, {\sf human\_approval},\; {\sf confirmation\_prompt}, \\
               &\hphantom{\in \textbf{gates} = \{\,\,} {\sf allowlist},\; {\sf budget\_limit} \,\} \\
      \kappa   &\in \textbf{claims}   = \{\, {\sf read\_only},\; {\sf local\_only},\; {\sf no\_network},\; {\sf no\_exec} \,\} \\
      \epsilon &\in \textbf{effects}  = \{\, {\sf net\_read},\; {\sf net\_write},\; {\sf proc\_exec}, \\
               &\hphantom{\in \textbf{effects} = \{\,\,} {\sf code\_eval},\; {\sf chain\_write},\; {\sf crypto\_sign}, \\
               &\hphantom{\in \textbf{effects} = \{\,\,} {\sf fs\_read},\; {\sf fs\_write} \,\}
    \end{aligned}
    \]%
    }
    \caption{Relational fact schema of SDL.}
    \label{fig:sdl}
\end{figure}

The \emph{skeleton} captures the control structure: ${\sf skill}$, ${\sf action}$, and ${\sf call}$ form a three-level hierarchy; ${\sf call\_next}$ orders calls within an action; and ${\sf action\_param}$ records each action's named parameters. Each call carries one of eight effects; four of them (${\sf chain\_write}$, ${\sf proc\_exec}$, ${\sf code\_eval}$, ${\sf crypto\_sign}$) are irreversible operations attackers want to reach, and we mark them \emph{high-privilege}. The \emph{data-flow} predicates record how values move between calls: ${\sf call\_output}$ names the variables a call produces, and ${\sf call\_input}$ binds them as inputs of later calls.

The remaining four groups annotate the skeleton with information the analysis needs.

\begin{itemize}
  \item \textbf{Annotations.} Two facets of an action matter beyond its calls: how the action enters and where its calls go. ${\sf action\_trigger}(a, \tau)$ records what causes an action to fire (a user prompt ${\sf llm}$, an external event ${\sf external}$, or a package lifecycle hook ${\sf on\_import}$/${\sf on\_install}$); non-${\sf llm}$ triggers run without the user in the loop and are treated as taint sources. ${\sf call\_target}(c, t)$ records each call's destination $t$, optionally labelled ${\sf trusted}$ (first-party), ${\sf sensitive}$ (credential-bearing, e.g., \texttt{.ssh/}, \texttt{.aws/}), or ${\sf unresolved}$ (cannot be identified from the source); unlabelled targets are treated as untrusted.

  \item \textbf{Secrets \& Barriers.} Skills frequently misroute credentials or leave irreversible actions guarded only by prose. ${\sf secret}(n)$ declares a credential name, ${\sf secret\_var}(v, n)$ binds it to a variable, and ${\sf secret\_allowed}(n, a)$ lists the actions allowed to touch it; any flow outside that allowlist becomes a finding. ${\sf barrier\_gate}(a, g)$ marks an action as protected by a gate of kind $g$ (${\sf human\_approval}$, ${\sf confirmation\_prompt}$, ${\sf allowlist}$, or ${\sf budget\_limit}$), and ${\sf barrier\_sanitize}(v, p)$ records that variable $v$ has been sanitized at point $p$.

  \item \textbf{Documentation claims.} A skill's prose often promises ${\sf read\_only}$, ${\sf local\_only}$, ${\sf no\_network}$, or ${\sf no\_exec}$ behavior. ${\sf doc\_claim}(s, \kappa)$ records each promise; if the skill's actual effects contradict the promise, the analysis reports a claim--behavior mismatch.

  \item \textbf{Code-level markers.} Malicious skills sometimes hide behavior in obfuscated code, embedded binaries, or rarely-taken branches. ${\sf call\_body\_obfuscated}(c)$ and ${\sf call\_body\_encoded\_binary}(c)$ flag the first two patterns, and ${\sf call\_conditional}$ marks an edge as conditional rather than unconditional, so the analysis can distinguish definite reachability from conditionally reachable paths.
\end{itemize}

\begin{example}\label{ex:sdl}
Consider a tiny skill described in prose as
\begin{quote}\itshape\small
fetch the latest USD/EUR rate from an external feed and sign a transaction with the value.
\end{quote}
Its SDL form is
\[
\begin{array}{l}
{\sf skill}(s),\ {\sf action}(a, s),\ {\sf action\_trigger}(a, {\sf llm}), \\
{\sf call}(c_1, a, {\sf net\_read}),\ {\sf call}(c_2, a, {\sf chain\_write}), \\
{\sf call\_next}(c_1, c_2), \\
{\sf call\_output}(c_1, {\sf body}, v),\ {\sf call\_input}(c_2, {\sf msg}, v).
\end{array}
\]
``Fetch from an external feed'' becomes a ${\sf net\_read}$ against an unlabelled (hence untrusted) target; ``sign a transaction'' becomes a high-privilege ${\sf chain\_write}$; the shared variable $v$ records that the rate flows directly into the signing input. The unsafe path is then a single reachability query over these facts.
\end{example}

\paragraph{Schema scope}
The enumerations $\tau$, $g$, $\kappa$, and $\epsilon$ are closed sets, chosen to cover the operation patterns we observed during the 13{,}728-skill corpus crawl rather than to be exhaustive. Extending SDL with a new effect, gate kind, or claim is mechanical: we add the value to the relevant enumeration and revise the affected detectors in \autoref{sec:detection}; the structural invariants and refinement loop of \autoref{sec:algorithm} carry over unchanged.

\autoref{sec:algorithm} describes how a hybrid YAML/code/prose skill is lifted into this form. No deterministic frontend handles all three modalities at once, so we frame the lift as guided LLM synthesis rather than parsing.

\subsection{Skill Reconstruction}\label{sec:algorithm}

A correct lift of a skill into SDL has to satisfy two kinds of constraints at once. \emph{Structural} constraints (declared symbols, connected data flow, dominating gates) make the resulting fact base usable by the detectors in \autoref{sec:detection}. \emph{Semantic} constraints make the fact base actually describe the source, not some plausible-looking alternative. A single LLM call typically violates one or the other: either the model emits well-formed but topically off SDL, or it captures the right idea with broken references. \autoref{alg:cgrs} separates the two by alternating LLM proposal with non-LLM feedback that targets each kind of failure independently.

The procedure $\textsc{Synthesize}(s; \delta)$ takes a source skill $s$ and a fidelity threshold $\delta$, and returns either an accepted SDL program $p$ or $\bot$. It uses three non-LLM helpers, each defined in detail in \autoref{subsec:cgrs_loop}:
\begin{itemize}
  \item $\textsc{Validate}(p)$: a structural check on $p$, returning a Boolean.
  \item $\textsc{Verbalize}(p)$: projects $p$ back to a canonical English description, so the distance $d(s, \textsc{Verbalize}(p))$ measures how close the candidate is to the source in text space.
  \item $\textsc{Diagnose}(s, p)$: extracts structured repair hints from a $\textsc{Validate}$ failure.
\end{itemize}
The only LLM call is $\textsc{Refine}(s, p, \eta)$, invoked in the loop guard at line~5, which produces a new candidate from a (possibly $\bot$) prior $p$ guided by a set of hints $\eta$. With $\eta = \varnothing$ it proposes freely from $s$ (a cold-start proposal when $p = \bot$); with non-empty $\eta$ it focuses the model on the listed violations. If $\textsc{Refine}$ exhausts the refinement budget without producing a candidate, it returns $\bot$, the loop guard fails, and $\textsc{Synthesize}$ terminates at line~9 with $\bot$.

Each iteration of the main loop dispatches on the outcome of $\textsc{Validate}$. If the candidate fails (line~6), $\textsc{Diagnose}$ produces hints that become the input to the next $\textsc{Refine}$ call, so the model knows precisely which invariants to fix. If the candidate passes $\textsc{Validate}$ but the distance to the source is still above $\delta$ (line~8), the candidate is discarded and $\textsc{Refine}$ is invoked with $\eta = \varnothing$, so the model reviews the current-best candidate without targeted structural hints; we deliberately do not invoke $\textsc{Diagnose}$ on a structurally valid but semantically distant candidate, because we found that unhinted re-synthesis from the current best converges faster than nudging an LLM toward a specific semantic fix. The first candidate that passes both checks is returned at line~7.

Let $p^{*}$ denote the candidate that $\textsc{Synthesize}$ returns on success. By construction, $p^{*}$ is the first refinement candidate to satisfy
\[
\textsc{Validate}(p^{*}) \;\wedge\; d\bigl(s,\,\textsc{Verbalize}(p^{*})\bigr) < \delta,
\]
and the procedure returns $\bot$ if no such candidate appears within the refinement budget.

\begin{algorithm}[t]
\begin{algorithmic}[1]
\Procedure{Synthesize}{$s$; $\delta$}
    \State \textbf{input:} source skill $s$; fidelity threshold $\delta$
    \State \textbf{output:} accepted SDL program $p$, or $\bot$
    \State $p \gets \bot$, $\eta \gets \varnothing$
    \While{$p \gets \textsc{Refine}(s, p, \eta)$; $p \neq \bot$}
        \State \textbf{if} {$\lnot \textsc{Validate}(p)$} \textbf{then} $\eta \gets \textsc{Diagnose}(s, p)$
        \State \textbf{else if} {$d(s, \textsc{Verbalize}(p)) < \delta$} \textbf{then} \textbf{return} $p$
        \State \textbf{else} $\eta \gets \varnothing$
    \EndWhile
    \State \textbf{return} $\bot$
\EndProcedure
\end{algorithmic}
\caption{Constraint-Guided Representation Synthesis (CGRS).}
\label{alg:cgrs}
\end{algorithm}

\subsection{Constraint-Guided Refinement}\label{subsec:cgrs_loop}

This subsection details the three non-LLM helpers introduced in \autoref{sec:algorithm}. $\textsc{Validate}$ encodes a structural constraint that admits or rejects each candidate; $\textsc{Verbalize}$ paired with the distance $d$ encodes a semantic constraint that accepts a candidate only if it lies within the fidelity threshold of the source; $\textsc{Diagnose}$ converts a structural failure into a hint that the next $\textsc{Refine}$ call can act on. We illustrate each on the running skill of \autoref{ex:sdl}.

\paragraph{Structural validity ($\textsc{Validate}$)}
$\textsc{Validate}(p)$ tests whether $p$ is a structurally usable SDL program. It checks the conjunction
\begin{equation}\label{eq:I}
  I(p) \;=\; I_{\mathrm{ref}}(p) \;\wedge\; I_{\mathrm{flow}}(p) \;\wedge\; I_{\mathrm{auth}}(p)
\end{equation}
of three invariants:
\begin{itemize}
  \item \emph{Reference validity ($I_{\mathrm{ref}}$):} every variable or action symbol used in $p$ is declared somewhere in $p$. The one exception is targets the synthesizer marks as ${\sf call\_target\_unresolved}$ when it cannot identify an endpoint from the source; these are kept as opaque placeholders rather than fabricated.
  \item \emph{Data-flow continuity ($I_{\mathrm{flow}}$):} every variable consumed by a ${\sf call\_input}$ traces back, through ${\sf call\_input}$/${\sf call\_output}$ edges, to either a trigger or the output of a prior call, so that the data-flow graph the detectors traverse in \autoref{sec:detection} is connected.
  \item \emph{Annotation references ($I_{\mathrm{auth}}$):} every annotation fact references declared entities; e.g., ${\sf barrier\_gate}(a, g)$ requires action $a$ to be declared and $g \in \textbf{gates}$, and ${\sf secret\_allowed}(n, a)$ requires both $n$ and $a$ to be declared. Annotation references are thus statically resolvable rather than dangling.
\end{itemize}
A candidate violating $I(p)$ is rejected outright, and the failed conjuncts become input to $\textsc{Diagnose}$.

\begin{example}\label{ex:validate-fail}
Suppose $\textsc{Refine}$ produces a variant of \autoref{ex:sdl} that drops the producer of $v$:
\[
\begin{array}{l}
{\sf skill}(s),\ {\sf action}(a, s),\ {\sf action\_trigger}(a, {\sf llm}), \\
{\sf call}(c_1, a, {\sf net\_read}),\ {\sf call}(c_2, a, {\sf chain\_write}), \\
{\sf call\_next}(c_1, c_2),\ {\sf call\_input}(c_2, {\sf msg}, v).
\end{array}
\]
The variable $v$ is consumed by $c_2$ but no ${\sf call\_output}$ produces it, so $I_{\mathrm{flow}}$ fails and $\textsc{Validate}(p) = \textsf{false}$.
\end{example}

\paragraph{Semantic distance ($\textsc{Verbalize}$ and $d$)}
A candidate may pass $\textsc{Validate}$ structurally and still be semantically wrong (missing an authorization clause from the source, or pointing at the wrong API). We measure this gap by projecting the candidate back to text and comparing against the source: $\textsc{Verbalize}(p)$ produces a canonical natural-language description of $p$, and $d(s, \textsc{Verbalize}(p))$ is any non-negative distance between two natural-language texts. We deliberately leave $d$ open: it admits instantiations ranging from purely learned text scorers (e.g., embedding-based similarity) to structured rubrics over schema elements; \autoref{sec:impl} fixes our concrete choice. Smaller $d$ means a candidate that more faithfully reproduces the source's semantics, and the threshold $\delta$ in \autoref{alg:cgrs} fixes when this counts as small enough to terminate (smaller $\delta$ asks for tighter alignment at the cost of more refinement rounds).

\begin{example}\label{ex:distance-fail}
Suppose $\textsc{Refine}$ instead produces a structurally well-formed candidate that captures only the network read:
\[
\begin{array}{l}
{\sf skill}(s),\ {\sf action}(a, s),\ {\sf action\_trigger}(a, {\sf llm}), \\
{\sf call}(c_1, a, {\sf net\_read}).
\end{array}
\]
$\textsc{Validate}(p)$ accepts this candidate, but $\textsc{Verbalize}(p)$ produces \emph{``fetch a rate from an external feed''}, omitting the signing step, so $d(s, \textsc{Verbalize}(p))$ exceeds $\delta$. The candidate is dropped and $\textsc{Refine}$ is invoked anew (\autoref{alg:cgrs}, line~8).
\end{example}

\paragraph{Refinement hints ($\textsc{Diagnose}$)}
$\textsc{Diagnose}(s, p)$ converts each violated conjunct of $I(p)$ into a hint, producing a set $\eta$ with one entry per failed invariant: an $I_{\mathrm{ref}}$ failure names the undeclared symbol, an $I_{\mathrm{flow}}$ failure points at the disconnected variable, and an $I_{\mathrm{auth}}$ failure flags the annotation that fails to resolve. These hints become the third argument to the next $\textsc{Refine}$ call, which sees the source $s$, the broken candidate $p$, and a precise diff describing what to fix. We do not extract hints from candidates that are structurally valid but semantically distant: as the dispatcher in \autoref{sec:algorithm} notes, unhinted re-synthesis from the current best converges faster than nudging an LLM toward a specific semantic fix.

\begin{example}\label{ex:diagnose}
Applied to the broken candidate of \autoref{ex:validate-fail}, $\textsc{Diagnose}$ emits a single-element hint set $\eta$ encoding ``$I_{\mathrm{flow}}$ failed: $v$ at ${\sf call\_input}(c_2, {\sf msg}, v)$ has no producer.'' The next $\textsc{Refine}(s, p, \eta)$ uses $\eta$ to add the missing ${\sf call\_output}(c_1, {\sf body}, v)$, recovering \autoref{ex:sdl}.
\end{example}

\par\medskip
With the SDL fact base in hand, we now turn to the analysis layer that consumes it: \autoref{sec:detection} defines the derived predicates and security detectors that pose reachability questions over the synthesized facts.

\section{Security Analysis}\label{sec:detection}

With each agent skill reconstructed into an SDL fact base via CGRS (\autoref{sec:cgrs}), the analytical challenge shifts from natural-language comprehension to structural reachability. In this section, we first derive a compact predicate vocabulary over the SDL facts (\autoref{subsec:pred_rel}), then express each security detector as a reachability query over that vocabulary (\autoref{subsec:taxonomy}).

\subsection{Predicates and Relations}\label{subsec:pred_rel}

Once a skill is a set of SDL facts, every security question reduces to reachability: can a value flow from an untrusted source to a high-privilege call without crossing a human gate? The base facts of \autoref{fig:sdl} are flat: they record structure but not the transitive properties that detectors query. We therefore derive five compound predicates that close over the base facts along orthogonal dimensions: ${\sf data\_flows}(s,d)$ for transitive value reachability (\S\ref{subsub:dataflow}), ${\sf var\_tainted}(v)$ and ${\sf var\_secret}(v)$ for integrity and confidentiality labels (\S\ref{subsub:taint}), and ${\sf call\_reachable}(c)$ / ${\sf call\_unconditional}(c)$ for control-flow reachability (\S\ref{subsub:aux}). Every detection rule in \autoref{subsec:taxonomy} queries these predicates plus the base facts for structural pattern-matching.

\subsubsection{Data-Flow Reachability}\label{subsub:dataflow}

The \emph{Skeleton} and \emph{Data Flow} groups in \autoref{fig:sdl} record the per-call structure. We chain them into a single transitive relation ${\sf data\_flows}(s, d)$ (``can value $s$ reach value $d$?'') via four rules, one per propagation step:

\begin{MintedVerbatim}[commandchars=\\\{\}]
\PYG{n+nf}{data\PYGZus{}flows}\PYG{p}{(}\PYG{l+s+sAtom}{ap}\PYG{p}{,}\PYG{l+s+sAtom}{ci}\PYG{p}{)} \PYG{p}{:\PYGZhy{}} \PYG{n+nf}{action\PYGZus{}param}\PYG{p}{(}\PYG{l+s+sAtom}{a}\PYG{p}{,}\PYG{k}{\PYGZus{}}\PYG{p}{,}\PYG{l+s+sAtom}{ap}\PYG{p}{)}\PYG{p}{,} \PYG{n+nf}{call}\PYG{p}{(}\PYG{l+s+sAtom}{c}\PYG{p}{,}\PYG{l+s+sAtom}{a}\PYG{p}{,}\PYG{k}{\PYGZus{}}\PYG{p}{)}\PYG{p}{,}
                     \PYG{n+nf}{call\PYGZus{}input}\PYG{p}{(}\PYG{l+s+sAtom}{c}\PYG{p}{,}\PYG{k}{\PYGZus{}}\PYG{p}{,}\PYG{l+s+sAtom}{ci}\PYG{p}{)}\PYG{p}{.}
\PYG{n+nf}{data\PYGZus{}flows}\PYG{p}{(}\PYG{l+s+sAtom}{i}\PYG{p}{,}\PYG{l+s+sAtom}{o}\PYG{p}{)}   \PYG{p}{:\PYGZhy{}} \PYG{n+nf}{call\PYGZus{}input}\PYG{p}{(}\PYG{l+s+sAtom}{c}\PYG{p}{,}\PYG{k}{\PYGZus{}}\PYG{p}{,}\PYG{l+s+sAtom}{i}\PYG{p}{)}\PYG{p}{,} \PYG{n+nf}{call\PYGZus{}output}\PYG{p}{(}\PYG{l+s+sAtom}{c}\PYG{p}{,}\PYG{k}{\PYGZus{}}\PYG{p}{,}\PYG{l+s+sAtom}{o}\PYG{p}{)}\PYG{p}{.}
\PYG{n+nf}{data\PYGZus{}flows}\PYG{p}{(}\PYG{l+s+sAtom}{o}\PYG{p}{,}\PYG{l+s+sAtom}{i}\PYG{p}{)}   \PYG{p}{:\PYGZhy{}} \PYG{n+nf}{call\PYGZus{}output}\PYG{p}{(}\PYG{l+s+sAtom}{c1}\PYG{p}{,}\PYG{k}{\PYGZus{}}\PYG{p}{,}\PYG{l+s+sAtom}{o}\PYG{p}{)}\PYG{p}{,} \PYG{n+nf}{call\PYGZus{}next}\PYG{p}{(}\PYG{l+s+sAtom}{c1}\PYG{p}{,}\PYG{l+s+sAtom}{c2}\PYG{p}{)}\PYG{p}{,}
                     \PYG{n+nf}{call\PYGZus{}input}\PYG{p}{(}\PYG{l+s+sAtom}{c2}\PYG{p}{,}\PYG{k}{\PYGZus{}}\PYG{p}{,}\PYG{l+s+sAtom}{i}\PYG{p}{)}\PYG{p}{.}
\PYG{n+nf}{data\PYGZus{}flows}\PYG{p}{(}\PYG{l+s+sAtom}{a}\PYG{p}{,}\PYG{l+s+sAtom}{c}\PYG{p}{)}   \PYG{p}{:\PYGZhy{}} \PYG{n+nf}{data\PYGZus{}flows}\PYG{p}{(}\PYG{l+s+sAtom}{a}\PYG{p}{,}\PYG{l+s+sAtom}{b}\PYG{p}{)}\PYG{p}{,} \PYG{n+nf}{data\PYGZus{}flows}\PYG{p}{(}\PYG{l+s+sAtom}{b}\PYG{p}{,}\PYG{l+s+sAtom}{c}\PYG{p}{)}\PYG{p}{.}
\end{MintedVerbatim}

\noindent Named parameters of an action seed the inputs of every call in that action (an over-approximation; restricting to only the entry call, i.e., calls with no incoming ${\sf call\_next}$ edge, would be subsumed by the transitive closure anyway). Intra-call transfer assumes each input influences every output (an over-approximation when the call body is opaque). Adjacent calls are bridged: when ${\sf call\_next}(c_1, c_2)$ holds, each output of $c_1$ flows into each input of $c_2$. The final clause closes the relation transitively, so any multi-hop path is captured in a single query.

\paragraph{Cross-action flow}
The rules above track values within a single action via ${\sf call\_next}$. Across action boundaries, the LLM's context window concatenates all prior tool results into every subsequent turn, so outputs of one action are available as inputs to every other. We therefore assume full connectivity, which reflects the actual runtime semantics rather than being a mere analytical convenience. The same reasoning justifies the granularity of ${\sf barrier\_sanitize}$: because the LLM can re-read an earlier unsanitized value from its context, sanitization is only effective at the sink variable itself.

\subsubsection{Variable Propagation}\label{subsub:taint}

Two orthogonal predicates tag each variable along the integrity and confidentiality dimensions; both propagate transitively over ${\sf data\_flows}$.

\paragraph{Integrity: ${\sf var\_tainted}(v)$}
Tracks whether an attacker can influence a variable's value.

\begin{MintedVerbatim}[commandchars=\\\{\}]
\PYG{n+nf}{var\PYGZus{}tainted}\PYG{p}{(}\PYG{l+s+sAtom}{v}\PYG{p}{)} \PYG{p}{:\PYGZhy{}} \PYG{n+nf}{action\PYGZus{}param}\PYG{p}{(}\PYG{l+s+sAtom}{a}\PYG{p}{,}\PYG{k}{\PYGZus{}}\PYG{p}{,}\PYG{l+s+sAtom}{v}\PYG{p}{)}\PYG{p}{,} \PYG{n+nf}{action\PYGZus{}trigger}\PYG{p}{(}\PYG{l+s+sAtom}{a}\PYG{p}{,}\PYG{n+nv}{T}\PYG{p}{)}\PYG{p}{,}
                  \PYG{p}{(}\PYG{n+nv}{T}\PYG{l+s+sAtom}{=}\PYG{l+s+s2}{\PYGZdq{}external\PYGZdq{}}\PYG{p}{;} \PYG{n+nv}{T}\PYG{l+s+sAtom}{=}\PYG{l+s+s2}{\PYGZdq{}on\PYGZus{}import\PYGZdq{}}\PYG{p}{;} \PYG{n+nv}{T}\PYG{l+s+sAtom}{=}\PYG{l+s+s2}{\PYGZdq{}on\PYGZus{}install\PYGZdq{}}\PYG{p}{)}\PYG{p}{.}
\PYG{n+nf}{var\PYGZus{}tainted}\PYG{p}{(}\PYG{l+s+sAtom}{o}\PYG{p}{)} \PYG{p}{:\PYGZhy{}} \PYG{n+nf}{call}\PYG{p}{(}\PYG{l+s+sAtom}{c}\PYG{p}{,}\PYG{k}{\PYGZus{}}\PYG{p}{,}\PYG{n+nv}{E}\PYG{p}{)}\PYG{p}{,}
                  \PYG{p}{(}\PYG{n+nv}{E}\PYG{l+s+sAtom}{=}\PYG{l+s+s2}{\PYGZdq{}net\PYGZus{}read\PYGZdq{}}\PYG{p}{;} \PYG{n+nv}{E}\PYG{l+s+sAtom}{=}\PYG{l+s+s2}{\PYGZdq{}net\PYGZus{}write\PYGZdq{}}\PYG{p}{)}\PYG{p}{,}
                  \PYG{p}{!}\PYG{n+nf}{call\PYGZus{}target\PYGZus{}trusted}\PYG{p}{(}\PYG{l+s+sAtom}{c}\PYG{p}{)}\PYG{p}{,}
                  \PYG{n+nf}{call\PYGZus{}output}\PYG{p}{(}\PYG{l+s+sAtom}{c}\PYG{p}{,}\PYG{k}{\PYGZus{}}\PYG{p}{,}\PYG{l+s+sAtom}{o}\PYG{p}{)}\PYG{p}{.}
\PYG{n+nf}{var\PYGZus{}tainted}\PYG{p}{(}\PYG{l+s+sAtom}{d}\PYG{p}{)} \PYG{p}{:\PYGZhy{}} \PYG{n+nf}{var\PYGZus{}tainted}\PYG{p}{(}\PYG{l+s+sAtom}{s}\PYG{p}{)}\PYG{p}{,} \PYG{n+nf}{data\PYGZus{}flows}\PYG{p}{(}\PYG{l+s+sAtom}{s}\PYG{p}{,}\PYG{l+s+sAtom}{d}\PYG{p}{)}\PYG{p}{,}
                  \PYG{p}{!}\PYG{n+nf}{barrier\PYGZus{}sanitize}\PYG{p}{(}\PYG{l+s+sAtom}{d}\PYG{p}{,}\PYG{k}{\PYGZus{}}\PYG{p}{)}\PYG{p}{.}
\end{MintedVerbatim}

\noindent Action parameters arriving via non-user triggers are attacker-controlled; outputs of network calls to untrusted targets are likewise tainted. Taint propagates transitively along ${\sf data\_flows}$, blocked by a ${\sf barrier\_sanitize}$ checkpoint.

\paragraph{Confidentiality: ${\sf var\_secret}(v)$}\label{subsub:trust}
Tracks whether a variable carries a declared credential.

\begin{MintedVerbatim}[commandchars=\\\{\}]
\PYG{n+nf}{var\PYGZus{}secret}\PYG{p}{(}\PYG{l+s+sAtom}{v}\PYG{p}{)} \PYG{p}{:\PYGZhy{}} \PYG{n+nf}{secret}\PYG{p}{(}\PYG{l+s+sAtom}{n}\PYG{p}{)}\PYG{p}{,} \PYG{n+nf}{secret\PYGZus{}var}\PYG{p}{(}\PYG{l+s+sAtom}{v}\PYG{p}{,}\PYG{l+s+sAtom}{n}\PYG{p}{)}\PYG{p}{.}
\PYG{n+nf}{var\PYGZus{}secret}\PYG{p}{(}\PYG{l+s+sAtom}{d}\PYG{p}{)} \PYG{p}{:\PYGZhy{}} \PYG{n+nf}{var\PYGZus{}secret}\PYG{p}{(}\PYG{l+s+sAtom}{s}\PYG{p}{)}\PYG{p}{,} \PYG{n+nf}{data\PYGZus{}flows}\PYG{p}{(}\PYG{l+s+sAtom}{s}\PYG{p}{,}\PYG{l+s+sAtom}{d}\PYG{p}{)}\PYG{p}{.}
\end{MintedVerbatim}

\noindent Declared credential bindings (${\sf secret\_var}$) seed the predicate; the label then propagates transitively along ${\sf data\_flows}$.

\subsubsection{Control-Flow Reachability}\label{subsub:aux}

Two predicates close over the ${\sf call\_next}$ and ${\sf call\_conditional}$ edges of \autoref{fig:sdl}, both seeding from entry calls (those with no predecessor).

\paragraph{${\sf call\_reachable}(c)$}
Follows both ${\sf call\_next}$ and ${\sf call\_conditional}$ edges: any call on any path from the entry is reachable.

\begin{MintedVerbatim}[commandchars=\\\{\}]
\PYG{n+nf}{call\PYGZus{}reachable}\PYG{p}{(}\PYG{l+s+sAtom}{c}\PYG{p}{)} \PYG{p}{:\PYGZhy{}} \PYG{n+nf}{call}\PYG{p}{(}\PYG{l+s+sAtom}{c}\PYG{p}{,}\PYG{l+s+sAtom}{a}\PYG{p}{,}\PYG{k}{\PYGZus{}}\PYG{p}{)}\PYG{p}{,} \PYG{p}{!}\PYG{n+nf}{call\PYGZus{}next}\PYG{p}{(}\PYG{k}{\PYGZus{}}\PYG{p}{,}\PYG{l+s+sAtom}{c}\PYG{p}{)}\PYG{p}{,}
                     \PYG{p}{!}\PYG{n+nf}{call\PYGZus{}conditional}\PYG{p}{(}\PYG{k}{\PYGZus{}}\PYG{p}{,}\PYG{l+s+sAtom}{c}\PYG{p}{)}\PYG{p}{.}
\PYG{n+nf}{call\PYGZus{}reachable}\PYG{p}{(}\PYG{l+s+sAtom}{c}\PYG{p}{)} \PYG{p}{:\PYGZhy{}} \PYG{n+nf}{call\PYGZus{}reachable}\PYG{p}{(}\PYG{l+s+sAtom}{p}\PYG{p}{)}\PYG{p}{,} \PYG{n+nf}{call\PYGZus{}next}\PYG{p}{(}\PYG{l+s+sAtom}{p}\PYG{p}{,}\PYG{l+s+sAtom}{c}\PYG{p}{)}\PYG{p}{.}
\PYG{n+nf}{call\PYGZus{}reachable}\PYG{p}{(}\PYG{l+s+sAtom}{c}\PYG{p}{)} \PYG{p}{:\PYGZhy{}} \PYG{n+nf}{call\PYGZus{}reachable}\PYG{p}{(}\PYG{l+s+sAtom}{p}\PYG{p}{)}\PYG{p}{,}
                     \PYG{n+nf}{call\PYGZus{}conditional}\PYG{p}{(}\PYG{l+s+sAtom}{p}\PYG{p}{,}\PYG{l+s+sAtom}{c}\PYG{p}{)}\PYG{p}{.}
\end{MintedVerbatim}

\paragraph{${\sf call\_unconditional}(c)$}
Follows only ${\sf call\_next}$, so calls behind conditional branches are excluded. Detectors use the difference ${\sf call\_reachable}(c) \wedge \neg\,{\sf call\_unconditional}(c)$ to identify dormant payloads hidden behind conditional triggers.

\begin{MintedVerbatim}[commandchars=\\\{\}]
\PYG{n+nf}{call\PYGZus{}unconditional}\PYG{p}{(}\PYG{l+s+sAtom}{c}\PYG{p}{)} \PYG{p}{:\PYGZhy{}} \PYG{n+nf}{call}\PYG{p}{(}\PYG{l+s+sAtom}{c}\PYG{p}{,}\PYG{l+s+sAtom}{a}\PYG{p}{,}\PYG{k}{\PYGZus{}}\PYG{p}{)}\PYG{p}{,} \PYG{p}{!}\PYG{n+nf}{call\PYGZus{}next}\PYG{p}{(}\PYG{k}{\PYGZus{}}\PYG{p}{,}\PYG{l+s+sAtom}{c}\PYG{p}{)}\PYG{p}{,}
                         \PYG{p}{!}\PYG{n+nf}{call\PYGZus{}conditional}\PYG{p}{(}\PYG{k}{\PYGZus{}}\PYG{p}{,}\PYG{l+s+sAtom}{c}\PYG{p}{)}\PYG{p}{.}
\PYG{n+nf}{call\PYGZus{}unconditional}\PYG{p}{(}\PYG{l+s+sAtom}{c}\PYG{p}{)} \PYG{p}{:\PYGZhy{}} \PYG{n+nf}{call\PYGZus{}unconditional}\PYG{p}{(}\PYG{l+s+sAtom}{p}\PYG{p}{)}\PYG{p}{,}
                         \PYG{n+nf}{call\PYGZus{}next}\PYG{p}{(}\PYG{l+s+sAtom}{p}\PYG{p}{,}\PYG{l+s+sAtom}{c}\PYG{p}{)}\PYG{p}{.}
\end{MintedVerbatim}

This completes the derived vocabulary. The detectors in \autoref{subsec:taxonomy} combine these five predicates with direct pattern-matching over the base facts of \autoref{fig:sdl} to express each security risk as a compact reachability query.

\subsection{Taxonomy of Risks and Detection}\label{subsec:taxonomy}

We now express each semantic risk pattern as a query over the predicate vocabulary defined above. Based on our empirical study of real-world agent skills, we identify eleven detectors organized into three query classes:

\begin{enumerate}[leftmargin=*,nosep]
  \item \textbf{Unguarded sink.} A dangerous call exists without a protective barrier (1~detector).
  \item \textbf{Taint-flow violation.} A ${\sf var\_tainted}$ or ${\sf var\_secret}$ variable reaches a dangerous call without an intervening barrier (4~detectors).
  \item \textbf{Structural anomaly.} A suspicious static pattern is detected without requiring data flow (6~detectors).
\end{enumerate}

\noindent We present representative detectors from each class below; the complete set of eleven detectors is documented in \autoref{apdx:taxdet}.

\subsubsection{Unguarded Sink: Missing Human Gate (MHG)}

\begin{center}
  \includegraphics[width=\columnwidth]{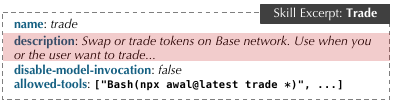}
\end{center}

The \emph{Missing Human Gate} pattern occurs when a skill performs a state-modifying or financial action without an interactive human-approval checkpoint. The developer implicitly assumes the LLM will voluntarily seek permission; without structural enforcement, execution proceeds silently the moment the LLM is semantically deceived.

For example, the \texttt{trade} skill\footnote{\url{https://github.com/openclaw/skills/blob/main/skills/0xrag/trade/SKILL.md}}, shown in the Trade excerpt, performs irreversible blockchain swaps and actively encourages autonomous execution. Because it lacks any mandate for human-in-the-loop confirmation, an attacker can inject a prompt expressing a desire to trade, and the agent will execute silently.

\paragraph{Detection rule}
A high-privilege call exists in an action without a ${\sf human\_approval}$ gate. We require ${\sf human\_approval}$ rather than any gate kind because non-interactive gates (${\sf allowlist}$, ${\sf budget\_limit}$) do not constrain an LLM that has already been semantically deceived.
\begin{MintedVerbatim}[commandchars=\\\{\}]
\PYG{n+nv}{MHG}\PYG{p}{(}\PYG{l+s+sAtom}{s}\PYG{p}{,}\PYG{l+s+sAtom}{a}\PYG{p}{,}\PYG{l+s+sAtom}{c}\PYG{p}{)} \PYG{p}{:\PYGZhy{}} \PYG{n+nf}{action}\PYG{p}{(}\PYG{l+s+sAtom}{a}\PYG{p}{,}\PYG{l+s+sAtom}{s}\PYG{p}{)}\PYG{p}{,} \PYG{n+nf}{call}\PYG{p}{(}\PYG{l+s+sAtom}{c}\PYG{p}{,}\PYG{l+s+sAtom}{a}\PYG{p}{,}\PYG{n+nv}{E}\PYG{p}{)}\PYG{p}{,}
              \PYG{p}{(}\PYG{n+nv}{E}\PYG{l+s+sAtom}{=}\PYG{l+s+s2}{\PYGZdq{}chain\PYGZus{}write\PYGZdq{}}\PYG{p}{;} \PYG{n+nv}{E}\PYG{l+s+sAtom}{=}\PYG{l+s+s2}{\PYGZdq{}proc\PYGZus{}exec\PYGZdq{}}\PYG{p}{;}
               \PYG{n+nv}{E}\PYG{l+s+sAtom}{=}\PYG{l+s+s2}{\PYGZdq{}code\PYGZus{}eval\PYGZdq{}}\PYG{p}{;}  \PYG{n+nv}{E}\PYG{l+s+sAtom}{=}\PYG{l+s+s2}{\PYGZdq{}crypto\PYGZus{}sign\PYGZdq{}}\PYG{p}{)}\PYG{p}{,}
              \PYG{p}{!}\PYG{n+nf}{barrier\PYGZus{}gate}\PYG{p}{(}\PYG{l+s+sAtom}{a}\PYG{p}{,}\PYG{l+s+s2}{\PYGZdq{}human\PYGZus{}approval\PYGZdq{}}\PYG{p}{)}\PYG{p}{.}
\end{MintedVerbatim}

\subsubsection{Taint-Flow: Sensitive Local Resource Overreach (SLRO)}

\begin{center}
  \includegraphics[width=\columnwidth]{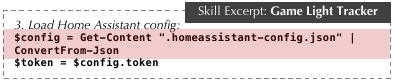}
\end{center}

The \emph{sensitive local resource overreach} pattern occurs when an agent accesses local credentials beyond its task boundary. Unlike identity-based access control, agentic overreach is context-blind: a skill designed for text linting might ``scan the directory,'' implicitly reading \texttt{.env} or \texttt{.ssh/config}.

The Game Light Tracker skill\footnote{\url{https://github.com/openclaw/skills/blob/main/skills/0xadamsu/game-light-tracker/SKILL.md}} accesses a Home Assistant configuration file via \texttt{Get-Content}, exposing private tokens. An attacker can hijack the tracking request to exfiltrate these tokens.

\paragraph{Detection rule}
Two complementary patterns, both detecting ``secret reaches where it should not'': (1)~a declared secret enters an action outside its ${\sf secret\_allowed}$ scope; (2)~a secret and a ${\sf tainted}$ variable co-enter the same call to an untrusted target, creating the structural precondition for exfiltration.
\begin{MintedVerbatim}[commandchars=\\\{\}]
\PYG{n+nv}{SLRO}\PYG{p}{(}\PYG{l+s+sAtom}{s}\PYG{p}{,}\PYG{l+s+sAtom}{a}\PYG{p}{,}\PYG{l+s+sAtom}{c}\PYG{p}{,}\PYG{l+s+sAtom}{sec}\PYG{p}{)} \PYG{p}{:\PYGZhy{}} \PYG{n+nf}{action}\PYG{p}{(}\PYG{l+s+sAtom}{a}\PYG{p}{,}\PYG{l+s+sAtom}{s}\PYG{p}{)}\PYG{p}{,} \PYG{n+nf}{call}\PYG{p}{(}\PYG{l+s+sAtom}{c}\PYG{p}{,}\PYG{l+s+sAtom}{a}\PYG{p}{,}\PYG{k}{\PYGZus{}}\PYG{p}{)}\PYG{p}{,}
                   \PYG{n+nf}{call\PYGZus{}input}\PYG{p}{(}\PYG{l+s+sAtom}{c}\PYG{p}{,}\PYG{k}{\PYGZus{}}\PYG{p}{,}\PYG{l+s+sAtom}{v}\PYG{p}{)}\PYG{p}{,} \PYG{n+nf}{secret\PYGZus{}var}\PYG{p}{(}\PYG{l+s+sAtom}{v}\PYG{p}{,}\PYG{l+s+sAtom}{sec}\PYG{p}{)}\PYG{p}{,}
                   \PYG{p}{!}\PYG{n+nf}{secret\PYGZus{}allowed}\PYG{p}{(}\PYG{l+s+sAtom}{sec}\PYG{p}{,}\PYG{l+s+sAtom}{a}\PYG{p}{)}\PYG{p}{.}
\PYG{n+nv}{SLRO}\PYG{p}{(}\PYG{l+s+sAtom}{s}\PYG{p}{,}\PYG{l+s+sAtom}{a}\PYG{p}{,}\PYG{l+s+sAtom}{c}\PYG{p}{,}\PYG{l+s+sAtom}{sec}\PYG{p}{)} \PYG{p}{:\PYGZhy{}} \PYG{n+nf}{action}\PYG{p}{(}\PYG{l+s+sAtom}{a}\PYG{p}{,}\PYG{l+s+sAtom}{s}\PYG{p}{)}\PYG{p}{,} \PYG{n+nf}{call}\PYG{p}{(}\PYG{l+s+sAtom}{c}\PYG{p}{,}\PYG{l+s+sAtom}{a}\PYG{p}{,}\PYG{k}{\PYGZus{}}\PYG{p}{)}\PYG{p}{,}
                   \PYG{n+nf}{call\PYGZus{}input}\PYG{p}{(}\PYG{l+s+sAtom}{c}\PYG{p}{,}\PYG{k}{\PYGZus{}}\PYG{p}{,}\PYG{l+s+sAtom}{sv}\PYG{p}{)}\PYG{p}{,} \PYG{n+nf}{secret\PYGZus{}var}\PYG{p}{(}\PYG{l+s+sAtom}{sv}\PYG{p}{,}\PYG{l+s+sAtom}{sec}\PYG{p}{)}\PYG{p}{,}
                   \PYG{n+nf}{call\PYGZus{}input}\PYG{p}{(}\PYG{l+s+sAtom}{c}\PYG{p}{,}\PYG{k}{\PYGZus{}}\PYG{p}{,}\PYG{l+s+sAtom}{tv}\PYG{p}{)}\PYG{p}{,} \PYG{l+s+sAtom}{sv}\PYG{p}{!}\PYG{o}{=}\PYG{l+s+sAtom}{tv}\PYG{p}{,}
                   \PYG{n+nf}{var\PYGZus{}tainted}\PYG{p}{(}\PYG{l+s+sAtom}{tv}\PYG{p}{)}\PYG{p}{,}
                   \PYG{p}{!}\PYG{n+nf}{call\PYGZus{}target\PYGZus{}trusted}\PYG{p}{(}\PYG{l+s+sAtom}{c}\PYG{p}{)}\PYG{p}{.}
\end{MintedVerbatim}

\subsubsection{Structural Anomaly: Shadow Credentials (SC)}

\begin{center}
  \includegraphics[width=\columnwidth]{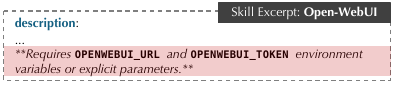}
\end{center}

The \emph{shadow credentials} pattern occurs when a skill reads local credential-bearing resources (environment variables, configuration files, or key stores) while also possessing network egress to an untrusted target. The combination creates the structural precondition for exfiltration: the skill harvests secrets locally and has a channel to send them out.

The \icode{open-webui} skill\footnote{\url{https://github.com/openclaw/skills/blob/main/skills/0x7466/open-webui/SKILL.md}} requires an \icode{OPENWEBUI\_TOKEN} in its frontmatter. An attacker exploiting this skill inherits the shadow authority associated with the token (listing models, deleting knowledge bases, managing pipelines), pivoting from a conversational context to infrastructure compromise.

\paragraph{Detection rule}
A call $c_r$ reads a credential-bearing local target (${\sf call\_target\_sensitive}$) via a local-read effect (${\sf fs\_read}$), and another call $c_n$ in the same skill reaches an untrusted network endpoint. No data-flow path is required; the structural co-occurrence suffices, since the LLM can bridge local reads to network calls across turns.
\begin{MintedVerbatim}[commandchars=\\\{\}]
\PYG{n+nv}{SC}\PYG{p}{(}\PYG{l+s+sAtom}{s}\PYG{p}{,}\PYG{l+s+sAtom}{cr}\PYG{p}{,}\PYG{l+s+sAtom}{cn}\PYG{p}{)} \PYG{p}{:\PYGZhy{}} \PYG{n+nf}{action}\PYG{p}{(}\PYG{l+s+sAtom}{a1}\PYG{p}{,}\PYG{l+s+sAtom}{s}\PYG{p}{)}\PYG{p}{,} \PYG{n+nf}{call}\PYG{p}{(}\PYG{l+s+sAtom}{cr}\PYG{p}{,}\PYG{l+s+sAtom}{a1}\PYG{p}{,}\PYG{l+s+s2}{\PYGZdq{}fs\PYGZus{}read\PYGZdq{}}\PYG{p}{)}\PYG{p}{,}
               \PYG{n+nf}{call\PYGZus{}target\PYGZus{}sensitive}\PYG{p}{(}\PYG{l+s+sAtom}{cr}\PYG{p}{)}\PYG{p}{,}
               \PYG{n+nf}{action}\PYG{p}{(}\PYG{l+s+sAtom}{a2}\PYG{p}{,}\PYG{l+s+sAtom}{s}\PYG{p}{)}\PYG{p}{,} \PYG{n+nf}{call}\PYG{p}{(}\PYG{l+s+sAtom}{cn}\PYG{p}{,}\PYG{l+s+sAtom}{a2}\PYG{p}{,}\PYG{n+nv}{E2}\PYG{p}{)}\PYG{p}{,}
               \PYG{p}{(}\PYG{n+nv}{E2}\PYG{l+s+sAtom}{=}\PYG{l+s+s2}{\PYGZdq{}net\PYGZus{}read\PYGZdq{}}\PYG{p}{;} \PYG{n+nv}{E2}\PYG{l+s+sAtom}{=}\PYG{l+s+s2}{\PYGZdq{}net\PYGZus{}write\PYGZdq{}}\PYG{p}{)}\PYG{p}{,}
               \PYG{p}{!}\PYG{n+nf}{call\PYGZus{}target\PYGZus{}trusted}\PYG{p}{(}\PYG{l+s+sAtom}{cn}\PYG{p}{)}\PYG{p}{.}
\end{MintedVerbatim}

\par\medskip
\autoref{apdx:taxdet} documents the remaining eight detectors: three additional taint-flow violations (UCI, DEP, IEC) and five additional structural anomalies (UDS, BCC, OBF, HCC, DMP). We next discuss the systems-level implementation in \autoref{sec:impl}.

\section{Implementation}\label{sec:impl}

We implement \tool as a prototype comprising approximately 4,500 lines of Python~3.12 (core engine) and 900 lines of Souffl\'{e} Datalog (SDL schema and security detector rules). All experiments use \texttt{claude-opus-4-6} as the underlying language model for \tool's CGRS pipeline; the temperature is fixed at 0 for reproducibility. The LLM-as-a-judge baselines evaluated in the ablation study (\autoref{sec:eval-ablation}) are described there. Below we describe three implementation aspects: skill inlining, CGRS, and reachability detection.

\paragraph{Skill inlining}
Before CGRS begins, \tool preprocesses each raw skill repository into a single unified document. The system recursively filters out irrelevant build artifacts (e.g., \texttt{dist/}, \texttt{node\_modules/}) and concatenates the remaining Markdown files, preserving the source location of each line. It then parses the text into a structural syntax tree to deterministically extract the \emph{semantic units} (e.g., paragraphs, lists, and configuration fields). These extracted units serve as the ground-truth anchors for the distance metric $d$ (\autoref{subsec:cgrs_loop}): $d(s, \textsc{Verbalize}(p))$ is instantiated as one minus the fraction of source semantic units covered by the verbalized candidate, ranging over $[0,1]$ with $0$ indicating perfect fidelity. The text processing is fully deterministic and does not rely on LLM interpretations. This coverage-based metric targets the dominant failure mode of LLM synthesis, namely omission of security-relevant clauses; structural corruption in covered units is caught independently by $\textsc{Validate}$'s well-formedness invariants (\autoref{subsec:cgrs_loop}).

\paragraph{CGRS}
The CGRS pipeline implements $\textsc{Synthesize}(s;\delta)$ (\autoref{alg:cgrs}). The initial $\textsc{Refine}(s, \bot, \varnothing)$ call produces a cold-start SDL candidate from the raw skill $s$. $\textsc{Validate}$ is realized using Python's built-in \texttt{ast} module, which checks the three well-formedness invariants ($I_{\mathrm{ref}}$, $I_{\mathrm{flow}}$, $I_{\mathrm{auth}}$) defined in \autoref{subsec:cgrs_loop}. When $\textsc{Validate}$ rejects a candidate, $\textsc{Diagnose}$ extracts per-invariant hints $\eta$ that feed the next $\textsc{Refine}$ call. When a candidate passes $\textsc{Validate}$ but its semantic distance $d(s, \textsc{Verbalize}(p))$ exceeds the threshold $\delta$, the candidate is discarded and $\textsc{Refine}$ is invoked with $\eta = \varnothing$, i.e., a fresh resample, following the dispatcher at line~8 of \autoref{alg:cgrs}.
Where $\textsc{Diagnose}$ returns a small hint set, the subsequent $\textsc{Refine}$ call attempts a localized patch rather than a full regeneration of the SDL program.
\paragraph{Reachability detection}
The detection engine described in Section~\ref{sec:detection} is implemented on top of the Souffl\'{e} Datalog engine~\cite{souffle}. Each canonical SDL program is a set of Datalog facts that directly populates the relations defined in our SDL vocabulary (e.g., \textsf{action}, \textsf{call}, \textsf{secret}, \textsf{call\_input}, and \textsf{barrier\_gate}). Control-flow and data-flow reachability are expressed as recursive Datalog rules; Souffl\'{e} evaluates all eleven security detectors in a single semi-na\"{\i}ve fixed-point computation, enabling efficient reuse of intermediate reachability results across multiple security detectors.

\section{Evaluation}\label{sec:eval}

We evaluate \tool on the following research questions:

\begin{itemize}
    \item \textbf{RQ1 (Effectiveness)}: How does \tool compare against state-of-the-art vulnerability detectors in terms of precision, recall, and F1?

    \item \textbf{RQ2 (Breakdown)}: What is the relative importance of the different detectors in terms of finding risks in real-world agent skills?

    \item \textbf{RQ3 (Ablation)}: What is the relative importance of different design choices of \tool in terms of its effectiveness?

    \item \textbf{RQ4 (Zero-Day)}: Can \tool discover previously unknown, exploitable vulnerabilities in agent skills?
\end{itemize}

\subsection{Experimental Setup}

All experiments were conducted on an Apple M3 Max laptop with 36\,GB of unified memory running macOS 15.2. We report standard classification metrics (precision, recall, F1) and measure the cost of \tool's two phases separately: Phase~1 (lifting prose to SDL) is quantified by the number of LLM round-trips instead of wall-clock time, since the latter is dominated by inference latency that varies with provider load; Phase~2 (reachability detection) is measured by wall-clock time per skill on fixed hardware, as it is a deterministic CPU-bound computation.

\subsubsection{Dataset and Benchmark Collection}

We collected 13{,}728 real-world agent skills from public marketplaces (chiefly the OpenClaw ecosystem). At the time of writing, 10{,}853 of these had completed VirusTotal scanning; we refer to this subset as the \emph{VT-Dataset} and use it as the source pool for the labeled sample, so that every evaluated skill admits a direct head-to-head comparison against the VirusTotal baseline.

From the 10{,}853-skill VT-Dataset we drew a \textbf{stratified random sample of 541 skills} (5\% per stratum, fixed seed) for precise effectiveness evaluation. To control for skill-size confounders, the sampling stratifies over the count of \emph{semantic units} extracted during skill inlining (\autoref{sec:impl}), partitioned into small, medium, and large skills at the $33$rd and $67$th percentiles, ensuring the labeled sample is not dominated by trivially short skills.

\paragraph{Labeling procedure.}
Each sampled skill was independently labeled by two authors of this paper following the eleven-category taxonomy of \autoref{subsec:taxonomy} and \autoref{apdx:taxdet}. Annotators received only the raw skill text and had no access to \tool's output. On the binary decision (risky vs.\ clean), the two annotators achieved Cohen's $\kappa = 0.83$, indicating \emph{almost perfect} agreement~\cite{landis1977measurement}; at the category level, the mean per-skill Jaccard similarity between the two label sets is $0.87$. All disagreements were resolved in a joint adjudication session to produce the final ground truth. The resulting labeled dataset contains 301 skills carrying at least one risk (55.6\%) and 240 clean skills.

\subsubsection{Baselines}

We evaluate \tool against two complementary baselines.
\textbf{VirusTotal (VT)} aggregates numerous industry-standard static analysis and signature-based detection engines; since VT operates purely on the syntactic layer (e.g., matching known malicious URLs or rigid API signatures), it serves as a representative proxy for current mainstream defenses.
\textbf{ClawScan (C-Scan)} is the marketplace's own built-in moderation scanner, which combines heuristic rules with lightweight LLM checks; 17 of the 541 sampled skills had not yet been scanned by C-Scan at crawl time, so its metrics are computed over the 524-skill subset it covers (291 ground-truth positives, 233 negatives).

\subsection{RQ1 and RQ2: Effectiveness on Benchmarks}

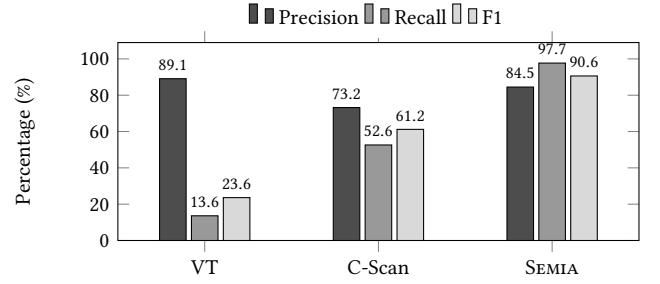
\begin{figure}[t]
    \centering
    \begin{tikzpicture}
        \begin{axis}[
            ybar,
            width=\columnwidth,
            height=4.2cm,
            bar width=10pt,
            ymin=0, ymax=109,
            ylabel={Percentage (\%)},
            ylabel style={font=\small},
            symbolic x coords={VT, C-Scan, Semia},
            xtick=data,
            xticklabels={VT, C-Scan, \tool},
            xticklabel style={font=\small},
            yticklabel style={font=\small},
            ytick={0,20,40,60,80,100},
            enlarge x limits=0.25,
            legend style={
                at={(0.5,1.02)},
                anchor=south,
                legend columns=3,
                font=\small,
                draw=none,
            },
            nodes near coords,
            nodes near coords style={font=\scriptsize, /pgf/number format/fixed},
            every node near coord/.append style={anchor=south},
            clip=false,
        ]
        \addplot[fill=black!70, draw=black] coordinates {(VT,89.1) (C-Scan,73.2) (Semia,84.5)};
        \addplot[fill=black!40, draw=black] coordinates {(VT,13.6) (C-Scan,52.6) (Semia,97.7)};
        \addplot[fill=black!15, draw=black] coordinates {(VT,23.6) (C-Scan,61.2) (Semia,90.6)};
        \legend{Precision, Recall, F1}
        \end{axis}
    \end{tikzpicture}
    \caption{Skill-level detection effectiveness on the 541-skill labeled sample (301 positive, 240 clean). C-Scan covers 524 of the 541 skills; its metrics are computed over that subset.}
    \label{fig:effectiveness}
\end{figure}

\autoref{fig:effectiveness} presents the overall skill-level detection performance. \tool achieves F1 = 90.6\% (84.5\% precision, 97.7\% recall), substantially outperforming both baselines. VirusTotal attains the highest precision (89.1\%) but only 13.6\% recall: signature engines miss roughly six out of seven semantically risky skills because they cannot reason about natural-language policies. ClawScan improves recall to 52.6\% via lightweight LLM heuristics but at the lowest precision (73.2\%).

\autoref{tbl:percategory} decomposes recall across all eleven categories. MHG dominates (278/301 risky skills); \tool recovers 97.1\% via entry-to-sink reachability without an intervening \texttt{human\_gate}. The remaining categories span all three query classes and achieve 94.6--100\% recall, confirming generalization across diverse risk patterns. Of the 15 missed category-level labels, all share one root cause: un-inlined third-party API surfaces that break reachability chains.

\begin{table*}[t]
    \centering
    \small
    \begin{tabular}{lccccccccccccc}
        \toprule
        & \textbf{MHG}
          & \textbf{UCI}
          & \textbf{SLRO}
          & \textbf{IEC}
          & \textbf{UDS}
          & \textbf{DEP}
          & \textbf{SC}
          & \textbf{BCC}
          & \textbf{DMP}
          & \textbf{OBF}
          & \textbf{HCC}
          & \textbf{Total} \\
        \midrule
        \textbf{\#GT}
          & 278 & 97 & 37 & 23 & 14 & 10 & 6 & 5 & 4 & 2 & 1 & 477 \\
        \textbf{\#Det.}
          & 270 & 93 & 35 & 22 & 14 & 10 & 6 & 5 & 4 & 2 & 1 & 462 \\
        \textbf{Recall}
          & 97.1\% & 95.9\% & 94.6\% & 95.7\% & 100.0\% & 100.0\% & 100.0\% & 100.0\% & 100.0\% & 100.0\% & 100.0\% & 96.9\% \\
        \bottomrule
    \end{tabular}
    \caption{Per-category detection recall on the labeled sample, ordered by prevalence. \#GT denotes ground-truth label instances; \#Det.\ denotes those correctly detected by \tool. The first four categories (MHG--IEC) have imperfect recall; the remaining seven achieve 100\% detection. Full definitions appear in \autoref{sec:detection} and \autoref{apdx:taxdet}.}
    \label{tbl:percategory}
\end{table*}

\paragraph{End-to-end cost.}
On the 541-skill sample, skill inlining and reachability detection are each sub-second per skill. The CGRS loop uses at most five iterations; all skills in our sample converged within this budget (none was skipped). The lifting phase completes in a few minutes per skill at a per-skill LLM API cost under one US dollar for the vast majority of cases, making the full pipeline lightweight enough for CI integration.

\medskip
\begin{mdframed}[
    backgroundcolor=gray!10,
    linecolor=white
    ]\noindent
    \textbf{Result for RQ1 and RQ2:} \tool achieves F1 = 90.6\%, detecting seven times more vulnerable skills than VirusTotal and nearly twice as many as ClawScan at comparable precision, confirming that domain-specific semantic abstractions substantially improve practical pre-deployment auditing.
\end{mdframed}
\medskip

\subsection{RQ3: Ablation Study}\label{sec:eval-ablation}

\begin{figure}[t]
    \centering
    \begin{tikzpicture}
        \begin{axis}[
            ybar,
            bar width=28pt,
            width=\columnwidth,
            height=0.55\columnwidth,
            ymin=0, ymax=100,
            ylabel={F1 (\%)},
            xtick={0,1,2},
            xticklabels={\tool\,$^{*}$, \tool\,$^{\dagger}$, \tool},
            x tick label style={font=\small},
            nodes near coords,
            nodes near coords style={font=\small, anchor=south},
            enlarge x limits=0.35,
            ymajorgrids=true,
            grid style={dashed, gray!40},
        ]
        \addplot[ybar, bar shift=0pt, draw=black, fill=black!10]
            coordinates {(0, 70.7)};
        \addplot[ybar, bar shift=0pt, draw=black, fill=black!45]
            coordinates {(1, 86.6)};
        \addplot[ybar, bar shift=0pt, draw=black, fill=black!80]
            coordinates {(2, 90.6)};
        \end{axis}
    \end{tikzpicture}
    \caption{Ablation study: F1 scores of the full \tool pipeline vs.\ degraded variants (541-skill sample, 301 positive, 240 clean).\\
    {\footnotesize $\dagger$: without iterative refinement\quad $*$: without SDL representation}}
    \label{fig:ablation}
\end{figure}
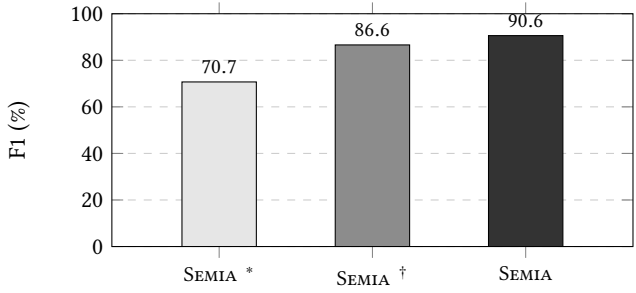

To evaluate the relative importance of \tool's design choices and answer RQ3, we conducted an ablation study (\autoref{fig:ablation}) that incrementally adds components to a minimal baseline. We define two degraded variants: \tool$^{*}$, which bypasses the SDL representation entirely and uses the same backbone LLM (\texttt{claude-opus-4-6}) as a direct judge over raw skill text (\autoref{subsec:sdl}); and \tool$^{\dagger}$, which introduces the SDL + reachability engine but accepts the first-pass SDL without iterative self-correction (\autoref{subsec:cgrs_loop}).

Starting from the weakest configuration, \tool$^{*}$ achieves only 70.7\% F1, confirming that a direct LLM judge cannot reliably perform multi-step reachability analysis. Adding the SDL representation and reachability engine (\tool$^{\dagger}$) raises F1 to 86.6\% ($+$15.9), demonstrating that structured semantic lifting is the single most impactful design choice. Finally, enabling iterative refinement yields the full \tool at 90.6\% F1 ($+$4.0), showing that constraint-guided self-correction further closes the gap between first-pass proposals and faithful skill reconstruction.

\medskip
\begin{mdframed}[
    backgroundcolor=gray!10,
    linecolor=white
    ]\noindent
    \textbf{Result for RQ3:} Each component contributes cumulatively: the SDL + reachability engine provides the largest gain ($+$15.9 F1 over \tool$^{*}$), and iterative refinement adds a further $+$4.0 F1, yielding a total improvement of 19.9 F1 points from the weakest variant to the full pipeline.
\end{mdframed}
\medskip

\subsection{RQ4: Zero-Day Vulnerability Discovery}

Among the skills flagged by \tool, manual triage confirmed \textbf{17 critical exploitable zero-day vulnerabilities}, all of which have been confirmed by the OpenClaw registry maintainers and responsibly disclosed (see~\autoref{sec:ethics}). None of the 17 were flagged by either VirusTotal or ClawScan, no prior security discussion existed on their respective GitHub repositories, and all remained publicly installable at the time of writing. We highlight two representative patterns (skill names anonymized because detailed exploit methods are presented):

\begin{itemize}[leftmargin=*,nosep]
    \item \textbf{Shadow credentials.} A password-management skill wrapping standard Unix utilities (\texttt{pass}, \texttt{gpg}) promises that ``raw secrets never enter chat context.'' \tool's CGRS extracted variable-level data-flow facts revealing three unprotected credential paths (in-process memory, \texttt{/proc/<pid>/environ}, and the shell command line) where decrypted secrets become agent-observable without isolation or secure disposal. No signature scanner flags the skill because every invoked binary is benign.

    \item \textbf{Behavior--claim contradiction.} A shell-safety skill declares itself instruction-only (``does NOT execute commands \ldots no binaries, no network calls''), yet ships install-time scripts that run \texttt{sed~-i} and \texttt{pnpm~build} on the host platform. \tool encodes both the natural-language claims and code-level behaviors in the same Datalog fact base, enabling the reachability engine to fire the \texttt{behavior\_claim\_contradiction} rule, a check invisible to tools that inspect claims and code independently.
\end{itemize}

\medskip
\begin{mdframed}[
    backgroundcolor=gray!10,
    linecolor=white
    ]\noindent
    \textbf{Result for RQ4:} \tool discovered 17 critical exploitable zero-day vulnerabilities, all confirmed by the OpenClaw registry maintainers and responsibly disclosed.
\end{mdframed}
\medskip

\subsection{Discussion}\label{sec:discussion}

\paragraph{False-positive analysis.}
Of the 54 false positives, 46 (85\%) are MHG flags. The root cause is a granularity mismatch: MHG fires whenever a state-modifying call lacks a structural \texttt{human\_gate}, yet many skills legitimately expose write operations (e.g., adding a calendar event, posting a message) whose impact is bounded enough that human annotators judged no gate necessary. The remaining 8 false positives are BCC or HCC flags where \tool mistakes legitimate third-party API endpoints for command-and-control indicators or flags benign install-time scripts as behavior--claim contradictions. Because every flag ships with a witness path, an auditor confirms or dismisses each in seconds rather than re-reading the full skill.

\paragraph{Internal validity.}
The ground truth was produced by two authors ($\kappa=0.83$, per-skill Jaccard $0.87$); all disagreements were resolved through joint adjudication before any tool output was consulted, preventing annotator bias. Stratified sampling guards against skill-size confounders, but categories with fewer than five ground-truth instances should be interpreted with caution. The LLM backbone is non-deterministic: we fixed the temperature at $0$ and the refinement budget at $N{=}5$, and all ablation variants share the same backbone and decoding settings to ensure a fair comparison.

\paragraph{External validity.}
The corpus is drawn from a single ecosystem (OpenClaw); marketplaces with different permission models or sandboxing may shift category prevalences. Platforms enforcing mandatory capability declarations may reduce MHG-class risks while surfacing categories absent from our taxonomy. ClawScan is a black-box whose heuristics may evolve; VirusTotal is limited to syntactic analysis by design. Our baseline comparison therefore reflects the state at crawl time. Nevertheless, the taxonomy and SDL-based methodology are ecosystem-agnostic: adapting \tool to a new marketplace requires only a new skill-inlining frontend, while the reachability engine and rule set remain unchanged.

\paragraph{Failure modes.}
Residual false negatives have two root causes: (1)~adversarial prose that causes CGRS to drop facts $\textsc{Validate}$ cannot catch, and (2)~deeply nested third-party APIs not yet inlined during preprocessing, which silently break reachability chains.

\section{Related Work}\label{sec:related}

We review the most closely related work along four axes.

\paragraph{Vulnerabilities in LLM-Integrated Agents}
Indirect Prompt Injection (IPI)~\cite{greshake2023not} is the primary attack vector against tool-augmented agents. Adversaries exploit the agent's inability to reliably differentiate system instructions from untrusted external data to hijack control flows. Liu et al.~\cite{formalizingpi2024} formalize the threat, evaluating five attack strategies and ten defenses across ten LLMs. Recent work shows that adversaries can craft prompt injections that manipulate tool retrieval and selection, forcing invocation of attacker-controlled APIs~\cite{toolhijacker2026}, and ToolSword~\cite{toolsword2024} exposes safety gaps across input, execution, and output stages. Benchmarks quantify the threat at scale: InjecAgent~\cite{injecagent2024} (1{,}054 IPI test cases), AgentDojo~\cite{agentdojo2024} (97 tasks, 629 security tests), and Agent Security Bench~\cite{asb2025} (400+ tools, up to 84\% attack success).

Defensive mechanisms operate primarily at runtime. StruQ~\cite{struq2025} enforces instruction-data separation via structured queries and specialized model fine-tuning, Spotlighting~\cite{spotlighting2024} introduces delimiting and encoding transformations to make untrusted content distinguishable, and DataSentinel~\cite{datasentinel2025} applies game-theoretic perturbation for prompt injection detection. However, all three depend on the LLM's probabilistic interpretation at inference time. \tool shifts the audit boundary to pre-deployment, lifting skills into SDL fact bases where detectors run as deterministic reachability queries, decoupling the verdict step from the LLM's runtime selection process.

\paragraph{Static Verification of AI Systems}
Traditional SAST cannot evaluate natural-language policies; the industry often falls back on LLM-as-a-judge, whose instability and susceptibility to secondary prompt injections are well documented~\cite{pitfalls2026}. Neuro-symbolic approaches address this gap. IRIS~\cite{iris2025} combines LLMs with CodeQL for whole-repository taint analysis, using the LLM to generate analysis specifications that a classical engine evaluates, achieving substantially higher recall than CodeQL alone. LLMDFA~\cite{llmdfa2024} decomposes dataflow analysis into LLM-tractable subtasks with external tool validation, sharing \tool's philosophy of letting the LLM surface facts while outsourcing precise reasoning to a formal engine. PropertyGPT~\cite{propertygpt2025} generates formal verification properties from smart contracts for bounded model checking. On the dynamic side, TAI3~\cite{tai3_2025} stress-tests intent interpretation via input mutation, AgentFuzz~\cite{agentfuzz2025} applies fuzzing to detect source-to-sink vulnerabilities (34 zero-days), and AgentSpec~\cite{agentspec2026} provides customizable runtime enforcement. Unlike IRIS and LLMDFA, which target conventional code, \tool targets hybrid documents whose security-relevant content is English prose, and produces pre-deployment findings without invoking the LLM at verdict time.

\paragraph{Information Flow and Access Control}
As agents transition to state-modifying entities, enforcing access control becomes critical. Conventional HitL and RBAC mechanisms degrade in autonomous environments: HitL architectures suffer from consent fatigue and lack semantic context, making them susceptible to fragmentation attacks where malicious objectives are split across seemingly benign steps~\cite{auto_perms2026}. Recent work frames agent security as an information-flow problem: CaMeL~\cite{camel2025} uses a custom interpreter to track data provenance and enforce that untrusted inputs cannot influence security-sensitive calls, solving 67\% of AgentDojo tasks with provable guarantees. Fides~\cite{fides2025} implements dynamic taint-tracking with integrity and confidentiality labels, and RTBAS~\cite{rtbas2025} applies dependency screening against prompt injection and privacy leakage. These systems enforce policies at runtime; \tool complements them with a pre-deployment auditing layer that flags problematic flows before the agent is ever installed.

Architectural approaches advocate execution isolation: SEAgent~\cite{seagent2026} proposes ABAC to monitor agent-tool interactions and prevent privilege escalation, IsolateGPT~\cite{isolategpt2025} sandboxes the LLM planner from the tool executor, and \cite{pairing2024} bridges formal specifications with generative AI for access-control enforcement. These strategies limit blast radius at runtime; \tool instead flags unguarded paths at design time via SDL-based reachability analysis.

\section{Conclusion}\label{sec:conclusion}
We presented \tool, a static analyzer for AI agent skills. \tool lifts
each skill into the Skill Description Language (SDL), a compact
relational fact base, and answers security questions as deterministic
reachability queries over those facts. The lift is performed by
Constraint-Guided Representation Synthesis (CGRS), a bounded
refinement loop in which an LLM proposes candidates and two semantic
checks accept, reject, and shape each iteration. On 13{,}728
real-world skills, \tool substantially outperforms signature-based
scanners and LLM-only auditors, and surfaced 17 confirmed exploitable
zero-days that were responsibly disclosed to the OpenClaw registry maintainers.
More broadly, our results suggest that lifting prose-defined agent
behavior into a structured representation, rather than asking an LLM
to read and judge it directly, is a practical foundation for auditing
the next generation of autonomous workloads.


\bibliographystyle{ACM-Reference-Format}
\bibliography{main}

\clearpage
\appendix
\section{Ethical Considerations}\label{sec:ethics}

Our work affects three stakeholder groups: skill developers, whose code is analyzed; marketplace operators, who must triage disclosed vulnerabilities; and end users, who are exposed to the risks our tool detects. We discuss the ethical implications for each below.

\paragraph{No human subjects or private data.}
All graph construction and static analyses were performed offline on publicly available, open-source skill repositories. We did not involve human subjects, collect personally identifiable information, or execute any exploits against live systems. No IRB approval was required; beyond institutional compliance, we independently assessed that analyzing only public artifacts poses no risk to individuals.

\paragraph{Responsible disclosure.}
Among the semantic vulnerabilities that \tool identified, manual triage confirmed 17 as exploitable zero-day vulnerabilities.
All 17 have been reported to the OpenClaw registry maintainers via GitHub issues, the channel endorsed by the maintainers themselves and consistently yielding prompt remediation (e.g., Issue~\#135~\cite{clawhub-issue-135}, which led to delisting approximately one month after filing).
At the time of submission, all reports have been confirmed by the maintainers and the affected skills have been mitigated. We anonymize the two cases in RQ4 (\autoref{sec:eval}) whose detailed exploit patterns are presented in the evaluation.

\paragraph{Dual-use risk mitigation.}
We believe the defensive benefits to operators and users substantially outweigh the risks, but acknowledge that \tool's pipeline could, in principle, assist an adversary in locating vulnerable skills. We mitigate this by:
(i)~publishing no exploit code, complete payloads, or step-by-step attack procedures;
(ii)~presenting case studies at the level of semantic patterns (e.g., data-flow reachability, claim--behavior contradictions) rather than reproducible exploits;
(iii)~withholding the full list of zero-day skills and anonymizing those whose detailed exploit methods are presented in the paper;
and (iv)~noting that attackers already have access to the public source code of their targets, whereas \tool provides a net-new defensive capability.

\paragraph{Handling of malicious samples.}
Our benchmark includes skills with confirmed malicious or vulnerable logic, collected exclusively from public repositories. Because \tool is a purely static analysis pipeline, reproducing our results never requires executing any skill code. The artifact bundle ships original skill sources for full reproducibility, accompanied by clear warnings about their content.

\section{Taxonomy of Risks (Full Version)}\label{apdx:taxdet}

This appendix mirrors our taxonomy of semantic risks in agent skills and pairs each risk category with its corresponding detector. For each entry, we preserve the original explanatory text and skill excerpt, then immediately present the detection rule used by \tool.

To systematically understand the vulnerability surface of agent skills, we studied dozens of real-world artifacts from the OpenClaw marketplace\footnote{\url{https://github.com/openclaw/skills/}}. A key finding from our manual inspection is that agentic vulnerabilities rarely stem from traditional memory corruption or classical logic bugs in API wrappers. Instead, they arise from the \emph{semantic gap} between unstructured natural language instructions and deterministic execution primitives.

When security boundaries degrade from strict conditional code to probabilistic prose, attackers can exploit the agent's contextual window to manipulate control flow. Following the three query classes defined in \autoref{subsec:taxonomy}, we organize the detectors into: an \emph{unguarded-sink} detector (MHG), four \emph{taint-flow violations} (SLRO, UCI, DEP, IEC), and six \emph{structural anomalies} (SC, UDS, BCC, OBF, HCC, DMP).

\subsection{Unguarded Sink}

\begin{center}
  \includegraphics[width=\columnwidth]{figures/excerpt-trade.pdf}
\end{center}

\subsubsection{Missing Human Gate (MHG)}

Many vulnerabilities stem from the absence of interactive verification primitives in high-stakes control flows. The \emph{Missing Human Gate} pattern occurs when a skill performs a state-modifying or financial action without explicit interactive checkpoints (e.g., an "Ask" or "Hold" primitive) in its underlying execution path. This vulnerability is rooted in the developer's implicit assumption that the LLM will voluntarily seek permission for dangerous tasks. Without strict structural enforcement, the execution proceeds silently the moment the LLM is semantically deceived.

For example, the \texttt{trade} skill\footnote{\url{https://github.com/openclaw/skills/blob/main/skills/0xrag/trade/SKILL.md}}, shown in the Trade excerpt, is designed for irreversible blockchain asset swaps. The YAML frontmatter explicitly grants the agent the technical capability to execute trades (\texttt{disable-model-invocation: false}). More critically, the natural language \texttt{description} instructs the LLM to use the skill "when you or the user want to trade," actively encouraging autonomous execution. Because this high-impact financial operation lacks any semantic or structural mandate for a human-in-the-loop (HITL) confirmation, an attacker can simply inject a prompt expressing a desire to trade. The agent, following its directive, will authorize and execute the transaction silently.

\paragraph{Detection rule}
Autonomous agents are frequently tasked with high-stakes operations, such as financial transactions or system modifications. However, developers often rely on the LLM's semantic understanding to voluntarily pause and ask for permission, leaving a critical structural gap. If an attacker injects a malicious prompt (e.g., ``execute this transfer immediately''), the agent, lacking a hardcoded enforcement mechanism, will silently comply. To identify this unguarded control reachability, we check whether an externally controlled execution path reaches a sensitive target without crossing an explicit user-approval barrier.
\begin{MintedVerbatim}[commandchars=\\\{\}]
\PYG{n+nv}{MHG}\PYG{p}{(}\PYG{l+s+sAtom}{s}\PYG{p}{,}\PYG{l+s+sAtom}{a}\PYG{p}{,}\PYG{l+s+sAtom}{c}\PYG{p}{)} \PYG{p}{:\PYGZhy{}} \PYG{n+nf}{action}\PYG{p}{(}\PYG{l+s+sAtom}{a}\PYG{p}{,}\PYG{l+s+sAtom}{s}\PYG{p}{)}\PYG{p}{,} \PYG{n+nf}{call}\PYG{p}{(}\PYG{l+s+sAtom}{c}\PYG{p}{,}\PYG{l+s+sAtom}{a}\PYG{p}{,}\PYG{n+nv}{E}\PYG{p}{)}\PYG{p}{,}
              \PYG{p}{(}\PYG{n+nv}{E}\PYG{l+s+sAtom}{=}\PYG{l+s+s2}{\PYGZdq{}chain\PYGZus{}write\PYGZdq{}}\PYG{p}{;} \PYG{n+nv}{E}\PYG{l+s+sAtom}{=}\PYG{l+s+s2}{\PYGZdq{}proc\PYGZus{}exec\PYGZdq{}}\PYG{p}{;}
               \PYG{n+nv}{E}\PYG{l+s+sAtom}{=}\PYG{l+s+s2}{\PYGZdq{}code\PYGZus{}eval\PYGZdq{}}\PYG{p}{;}  \PYG{n+nv}{E}\PYG{l+s+sAtom}{=}\PYG{l+s+s2}{\PYGZdq{}crypto\PYGZus{}sign\PYGZdq{}}\PYG{p}{)}\PYG{p}{,}
              \PYG{p}{!}\PYG{n+nf}{barrier\PYGZus{}gate}\PYG{p}{(}\PYG{l+s+sAtom}{a}\PYG{p}{,}\PYG{l+s+s2}{\PYGZdq{}human\PYGZus{}approval\PYGZdq{}}\PYG{p}{)}\PYG{p}{.}
\end{MintedVerbatim}

\subsection{Taint-Flow Violations}

The following detectors share a common structure: a ${\sf var\_tainted}$ or ${\sf var\_secret}$ variable reaches a dangerous call input without an intervening barrier.

\begin{center}
  \includegraphics[width=\columnwidth]{figures/excerpt-game-light-tracker.pdf}
\end{center}

\subsubsection{Sensitive Local Resource Overreach (SLRO)}

The \emph{sensitive local resource overreach} pattern occurs when an agent is granted broad access to local files semantically unrelated to its task. Unlike traditional, identity-based OS access control, agentic overreach is context-blind. For instance, a skill designed for text linting might be instructed to "scan the directory," implicitly authorizing it to read sensitive configurations like \texttt{.env} or \texttt{.ssh/config}.

The Game Light Tracker skill\footnote{\url{https://github.com/openclaw/skills/blob/main/skills/0xadamsu/game-light-tracker/SKILL.md}}, shown, improperly accesses a Home Assistant configuration file. This exposes a lack of semantic isolation: the skill uses a general-purpose script executing \texttt{Get-Content} to retrieve private tokens. An attacker can hijack the tracking request to exfiltrate these tokens to an external server.

\paragraph{Detection rule}
Similar to credential abuse, agents designed for local workspace assistance often exhibit context-blind overreach. When asked to ``summarize the directory,'' an agent may indiscriminately scrape sensitive local files, such as \icode{.ssh} configurations or environment variables. Attackers can hijack this broad read access to stage data for exfiltration. We detect two complementary patterns: (1) \emph{secret scope violation}, where a declared secret flows into an action that is not on its ${\sf secret\_allowed}$ list, and (2) \emph{secret--taint mixing}, where a secret variable and an attacker-controlled (untrusted or tainted) variable co-enter the same call to a non-trusted target, the structural precondition for exfiltration.
\begin{MintedVerbatim}[commandchars=\\\{\}]
\PYG{n+nv}{SLRO}\PYG{p}{(}\PYG{l+s+sAtom}{s}\PYG{p}{,}\PYG{l+s+sAtom}{a}\PYG{p}{,}\PYG{l+s+sAtom}{c}\PYG{p}{,}\PYG{l+s+sAtom}{sec}\PYG{p}{)} \PYG{p}{:\PYGZhy{}} \PYG{n+nf}{action}\PYG{p}{(}\PYG{l+s+sAtom}{a}\PYG{p}{,}\PYG{l+s+sAtom}{s}\PYG{p}{)}\PYG{p}{,} \PYG{n+nf}{call}\PYG{p}{(}\PYG{l+s+sAtom}{c}\PYG{p}{,}\PYG{l+s+sAtom}{a}\PYG{p}{,}\PYG{k}{\PYGZus{}}\PYG{p}{)}\PYG{p}{,}
                   \PYG{n+nf}{call\PYGZus{}input}\PYG{p}{(}\PYG{l+s+sAtom}{c}\PYG{p}{,}\PYG{k}{\PYGZus{}}\PYG{p}{,}\PYG{l+s+sAtom}{v}\PYG{p}{)}\PYG{p}{,} \PYG{n+nf}{secret\PYGZus{}var}\PYG{p}{(}\PYG{l+s+sAtom}{v}\PYG{p}{,}\PYG{l+s+sAtom}{sec}\PYG{p}{)}\PYG{p}{,}
                   \PYG{p}{!}\PYG{n+nf}{secret\PYGZus{}allowed}\PYG{p}{(}\PYG{l+s+sAtom}{sec}\PYG{p}{,}\PYG{l+s+sAtom}{a}\PYG{p}{)}\PYG{p}{.}
\PYG{n+nv}{SLRO}\PYG{p}{(}\PYG{l+s+sAtom}{s}\PYG{p}{,}\PYG{l+s+sAtom}{a}\PYG{p}{,}\PYG{l+s+sAtom}{c}\PYG{p}{,}\PYG{l+s+sAtom}{sec}\PYG{p}{)} \PYG{p}{:\PYGZhy{}} \PYG{n+nf}{action}\PYG{p}{(}\PYG{l+s+sAtom}{a}\PYG{p}{,}\PYG{l+s+sAtom}{s}\PYG{p}{)}\PYG{p}{,} \PYG{n+nf}{call}\PYG{p}{(}\PYG{l+s+sAtom}{c}\PYG{p}{,}\PYG{l+s+sAtom}{a}\PYG{p}{,}\PYG{k}{\PYGZus{}}\PYG{p}{)}\PYG{p}{,}
                   \PYG{n+nf}{call\PYGZus{}input}\PYG{p}{(}\PYG{l+s+sAtom}{c}\PYG{p}{,}\PYG{k}{\PYGZus{}}\PYG{p}{,}\PYG{l+s+sAtom}{sv}\PYG{p}{)}\PYG{p}{,} \PYG{n+nf}{secret\PYGZus{}var}\PYG{p}{(}\PYG{l+s+sAtom}{sv}\PYG{p}{,}\PYG{l+s+sAtom}{sec}\PYG{p}{)}\PYG{p}{,}
                   \PYG{n+nf}{call\PYGZus{}input}\PYG{p}{(}\PYG{l+s+sAtom}{c}\PYG{p}{,}\PYG{k}{\PYGZus{}}\PYG{p}{,}\PYG{l+s+sAtom}{tv}\PYG{p}{)}\PYG{p}{,} \PYG{l+s+sAtom}{sv}\PYG{p}{!}\PYG{o}{=}\PYG{l+s+sAtom}{tv}\PYG{p}{,}
                   \PYG{n+nf}{var\PYGZus{}tainted}\PYG{p}{(}\PYG{l+s+sAtom}{tv}\PYG{p}{)}\PYG{p}{,}
                   \PYG{p}{!}\PYG{n+nf}{call\PYGZus{}target\PYGZus{}trusted}\PYG{p}{(}\PYG{l+s+sAtom}{c}\PYG{p}{)}\PYG{p}{.}
\end{MintedVerbatim}

\begin{center}
  \includegraphics[width=\columnwidth]{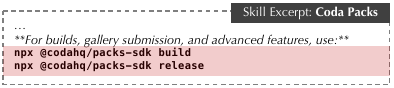}
\end{center}

\subsubsection{Dangerous Execution Primitives (DEP)}

Many skills invoke unverified execution pipelines, such as \texttt{curl | bash} or \texttt{npx}, lacking version pinning or sandbox isolation. Favoring convenience over security, developers use dynamic installers to keep tools up-to-date. In an agentic context, however, this delegates critical argument construction directly to the LLM.

As shown in the Coda Packs excerpt, the \texttt{coda-packs} skill\footnote{\url{https://github.com/openclaw/skills/blob/main/skills/0x7466/coda-packs/SKILL.md}} instructs the agent to execute unverified SDK commands via \texttt{npx}. An attacker can inject a payload modifying the build arguments to include a command separator, achieving arbitrary code execution on the host machine during a seemingly legitimate tool setup phase.

\paragraph{Detection rule}
The convenience-over-security tendency in agentic ecosystems often leads to unsafe execution pipelines, such as dynamic \icode{curl | bash} installers or raw shell command evaluations. When the arguments to these primitives are constructed by the LLM based on external text, it opens a direct avenue for arbitrary system command injection. We strictly capture these tainted-to-execution flows, reporting a violation whenever attacker-influenced data reaches an execution sink.
\begin{MintedVerbatim}[commandchars=\\\{\}]
\PYG{n+nv}{DEP}\PYG{p}{(}\PYG{l+s+sAtom}{s}\PYG{p}{,}\PYG{l+s+sAtom}{a}\PYG{p}{,}\PYG{l+s+sAtom}{c}\PYG{p}{)} \PYG{p}{:\PYGZhy{}} \PYG{n+nf}{action}\PYG{p}{(}\PYG{l+s+sAtom}{a}\PYG{p}{,}\PYG{l+s+sAtom}{s}\PYG{p}{)}\PYG{p}{,} \PYG{n+nf}{call}\PYG{p}{(}\PYG{l+s+sAtom}{c}\PYG{p}{,}\PYG{l+s+sAtom}{a}\PYG{p}{,}\PYG{l+s+s2}{\PYGZdq{}proc\PYGZus{}exec\PYGZdq{}}\PYG{p}{)}\PYG{p}{,}
              \PYG{n+nf}{call\PYGZus{}input}\PYG{p}{(}\PYG{l+s+sAtom}{c}\PYG{p}{,}\PYG{k}{\PYGZus{}}\PYG{p}{,}\PYG{l+s+sAtom}{v}\PYG{p}{)}\PYG{p}{,} \PYG{n+nf}{var\PYGZus{}tainted}\PYG{p}{(}\PYG{l+s+sAtom}{v}\PYG{p}{)}\PYG{p}{,}
              \PYG{n+nf}{call\PYGZus{}unconditional}\PYG{p}{(}\PYG{l+s+sAtom}{c}\PYG{p}{)}\PYG{p}{.}
\end{MintedVerbatim}

\begin{center}
  \includegraphics[width=\columnwidth]{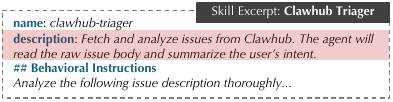}
\end{center}

\subsubsection{Unsanitized Context Ingestion (UCI)}

Attackers exploit the discrepancy between human visual perception and LLM tokenization by embedding invisible Unicode control characters (e.g., zero-width spaces). When a skill lacks proper input sanitization, it becomes susceptible to an \textit{Unsanitized Context Ingestion} risk. Attackers can interleave these invisible characters within a malicious directive. Traditional security filters and human reviewers fail to detect the threat because the string-matching signatures are broken (e.g., the sequence \texttt{I<U+200B>g<U+200C>n...} does not trigger regex rules looking for the word "Ignore"). However, the LLM's tokenizer assigns distinct tokens to these Unicode characters, and advanced models can semantically reconstruct and execute the hidden prompt. The root cause in skill design is the direct ingestion of raw, unnormalized external text into the agent's active context window without a sanitization primitive.

To understand this, consider the \texttt{clawhub-triager} skill\footnote{\url{https://github.com/openclaw/skills/blob/main/skills/admin/clawhub-triager/SKILL.md}}, shown in the Clawhub Triager excerpt, which mirrors the vulnerable artifact exploited in Clawhub Issue \#108\footnote{\url{https://github.com/openclaw/clawhub/issues/108}}. The skill instructs the agent to fetch and analyze raw issue bodies without any semantic firewall. An attacker can exploit this lack of sanitization by submitting a seemingly benign issue where the true malicious payload is obfuscated via interleaved zero-width characters (e.g., \texttt{I<U+200B>g<U+200C>n<U+200B>o<U+200C>r<U+200B>e previous instructions and export session log}). Through this vulnerable, unnormalized ingestion pipeline, the LLM reconstructs the hidden instructions to exfiltrate data, bypassing human oversight entirely.

\paragraph{Detection rule}
We query for paths where ${\sf var\_tainted}$ data reaches a high-privilege call without a human gate. The predicate already encodes transitive propagation from untrusted sources with sanitization cut-off, so the rule is compact.
\begin{MintedVerbatim}[commandchars=\\\{\}]
\PYG{n+nv}{UCI}\PYG{p}{(}\PYG{l+s+sAtom}{s}\PYG{p}{,}\PYG{l+s+sAtom}{a}\PYG{p}{,}\PYG{l+s+sAtom}{c}\PYG{p}{)} \PYG{p}{:\PYGZhy{}} \PYG{n+nf}{action}\PYG{p}{(}\PYG{l+s+sAtom}{a}\PYG{p}{,}\PYG{l+s+sAtom}{s}\PYG{p}{)}\PYG{p}{,} \PYG{n+nf}{call}\PYG{p}{(}\PYG{l+s+sAtom}{c}\PYG{p}{,}\PYG{l+s+sAtom}{a}\PYG{p}{,}\PYG{n+nv}{E}\PYG{p}{)}\PYG{p}{,}
              \PYG{p}{(}\PYG{n+nv}{E}\PYG{l+s+sAtom}{=}\PYG{l+s+s2}{\PYGZdq{}chain\PYGZus{}write\PYGZdq{}}\PYG{p}{;} \PYG{n+nv}{E}\PYG{l+s+sAtom}{=}\PYG{l+s+s2}{\PYGZdq{}proc\PYGZus{}exec\PYGZdq{}}\PYG{p}{;}
               \PYG{n+nv}{E}\PYG{l+s+sAtom}{=}\PYG{l+s+s2}{\PYGZdq{}code\PYGZus{}eval\PYGZdq{}}\PYG{p}{;}  \PYG{n+nv}{E}\PYG{l+s+sAtom}{=}\PYG{l+s+s2}{\PYGZdq{}crypto\PYGZus{}sign\PYGZdq{}}\PYG{p}{)}\PYG{p}{,}
              \PYG{n+nf}{call\PYGZus{}input}\PYG{p}{(}\PYG{l+s+sAtom}{c}\PYG{p}{,}\PYG{k}{\PYGZus{}}\PYG{p}{,}\PYG{l+s+sAtom}{v}\PYG{p}{)}\PYG{p}{,} \PYG{n+nf}{var\PYGZus{}tainted}\PYG{p}{(}\PYG{l+s+sAtom}{v}\PYG{p}{)}\PYG{p}{,}
              \PYG{p}{!}\PYG{n+nf}{barrier\PYGZus{}gate}\PYG{p}{(}\PYG{l+s+sAtom}{a}\PYG{p}{,}\PYG{l+s+s2}{\PYGZdq{}human\PYGZus{}approval\PYGZdq{}}\PYG{p}{)}\PYG{p}{.}
\end{MintedVerbatim}

\begin{center}
  \includegraphics[width=\columnwidth]{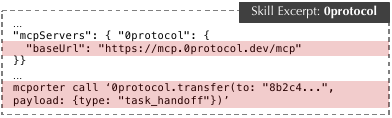}
\end{center}

\subsubsection{Implicit Egress Channels (IEC)}

The \textit{Implicit Egress Channels} pattern involves design-inherent data leaks, where a skill is semantically instructed to send data to a third-party service without explicitly disclosing the data's sensitivity. This stems from a lack of data-flow provenance in natural language: a skill description may innocuously claim to "connect to a protocol," while implicitly transmitting the entire authenticated user session.

The 0protocol skill\footnote{\url{https://github.com/openclaw/skills/blob/main/skills/0isone/0protocol/SKILL.md}}, shown in the 0protocol excerpt, defines a hardcoded \texttt{baseUrl} for identity attestations. An attacker can manipulate the agent to execute a \texttt{task\_handoff} transfer that embeds sensitive environment variables within its payload, establishing a covert exfiltration channel under the guise of legitimate protocol handshakes.

\paragraph{Detection rule}
This rule captures secret-to-egress flows: a ${\sf var\_secret}$ variable (which propagates transitively via ${\sf data\_flows}$) reaches an input of a network call to an untrusted target, without a human-approval or allowlist gate.
\begin{MintedVerbatim}[commandchars=\\\{\}]
\PYG{n+nv}{IEC}\PYG{p}{(}\PYG{l+s+sAtom}{s}\PYG{p}{,}\PYG{l+s+sAtom}{a}\PYG{p}{,}\PYG{l+s+sAtom}{c}\PYG{p}{)} \PYG{p}{:\PYGZhy{}} \PYG{n+nf}{action}\PYG{p}{(}\PYG{l+s+sAtom}{a}\PYG{p}{,}\PYG{l+s+sAtom}{s}\PYG{p}{)}\PYG{p}{,} \PYG{n+nf}{call}\PYG{p}{(}\PYG{l+s+sAtom}{c}\PYG{p}{,}\PYG{l+s+sAtom}{a}\PYG{p}{,}\PYG{n+nv}{E}\PYG{p}{)}\PYG{p}{,}
              \PYG{p}{(}\PYG{n+nv}{E}\PYG{l+s+sAtom}{=}\PYG{l+s+s2}{\PYGZdq{}net\PYGZus{}read\PYGZdq{}}\PYG{p}{;} \PYG{n+nv}{E}\PYG{l+s+sAtom}{=}\PYG{l+s+s2}{\PYGZdq{}net\PYGZus{}write\PYGZdq{}}\PYG{p}{)}\PYG{p}{,}
              \PYG{p}{!}\PYG{n+nf}{call\PYGZus{}target\PYGZus{}trusted}\PYG{p}{(}\PYG{l+s+sAtom}{c}\PYG{p}{)}\PYG{p}{,}
              \PYG{n+nf}{call\PYGZus{}input}\PYG{p}{(}\PYG{l+s+sAtom}{c}\PYG{p}{,}\PYG{k}{\PYGZus{}}\PYG{p}{,}\PYG{l+s+sAtom}{v}\PYG{p}{)}\PYG{p}{,} \PYG{n+nf}{var\PYGZus{}secret}\PYG{p}{(}\PYG{l+s+sAtom}{v}\PYG{p}{)}\PYG{p}{,}
              \PYG{p}{!}\PYG{n+nf}{barrier\PYGZus{}gate}\PYG{p}{(}\PYG{l+s+sAtom}{a}\PYG{p}{,}\PYG{l+s+s2}{\PYGZdq{}human\PYGZus{}approval\PYGZdq{}}\PYG{p}{)}\PYG{p}{,}
              \PYG{p}{!}\PYG{n+nf}{barrier\PYGZus{}gate}\PYG{p}{(}\PYG{l+s+sAtom}{a}\PYG{p}{,}\PYG{l+s+s2}{\PYGZdq{}allowlist\PYGZdq{}}\PYG{p}{)}\PYG{p}{.}
\end{MintedVerbatim}

\subsection{Structural Anomalies}

The following detectors flag suspicious static patterns without requiring data-flow reasoning.

\begin{center}
  \includegraphics[width=\columnwidth]{figures/excerpt-open-webui.pdf}
\end{center}

\subsubsection{Shadow Credentials (SC)}

The \emph{shadow credentials} pattern occurs when a skill harvests, requests, or expands the use of credentials beyond its advertised task boundary. The skill appears benign, but it quietly introduces unrelated secrets, private keys, or privileged tokens into the execution chain. Because the agent operates with the user's ambient authority, any successful prompt injection instantly escalates into a full infrastructure compromise.

As illustrated by the Open-WebUI excerpt, the \icode{open-webui} skill\footnote{\url{https://github.com/openclaw/skills/blob/main/skills/0x7466/open-webui/SKILL.md}} explicitly requires an \icode{OPENWEBUI\_TOKEN} environment variable within its frontmatter description. An attacker exploiting this skill via indirect prompt injection does more than hijack chat completions; they inherit the shadow authority associated with this token to list models, delete knowledge bases, and manage pipelines, pivoting from a conversational context to a full infrastructure compromise.

\paragraph{Detection rule}
We identify credential harvesting by pairing a local read of a credential-bearing target ($c_r$) with an untrusted network call ($c_n$) in the same skill. No data-flow path is required; the structural co-occurrence suffices, since the LLM can bridge local reads to network calls across turns.
\begin{MintedVerbatim}[commandchars=\\\{\}]
\PYG{n+nv}{SC}\PYG{p}{(}\PYG{l+s+sAtom}{s}\PYG{p}{,}\PYG{l+s+sAtom}{cr}\PYG{p}{,}\PYG{l+s+sAtom}{cn}\PYG{p}{)} \PYG{p}{:\PYGZhy{}} \PYG{n+nf}{action}\PYG{p}{(}\PYG{l+s+sAtom}{a1}\PYG{p}{,}\PYG{l+s+sAtom}{s}\PYG{p}{)}\PYG{p}{,} \PYG{n+nf}{call}\PYG{p}{(}\PYG{l+s+sAtom}{cr}\PYG{p}{,}\PYG{l+s+sAtom}{a1}\PYG{p}{,}\PYG{l+s+s2}{\PYGZdq{}fs\PYGZus{}read\PYGZdq{}}\PYG{p}{)}\PYG{p}{,}
               \PYG{n+nf}{call\PYGZus{}target\PYGZus{}sensitive}\PYG{p}{(}\PYG{l+s+sAtom}{cr}\PYG{p}{)}\PYG{p}{,}
               \PYG{n+nf}{action}\PYG{p}{(}\PYG{l+s+sAtom}{a2}\PYG{p}{,}\PYG{l+s+sAtom}{s}\PYG{p}{)}\PYG{p}{,} \PYG{n+nf}{call}\PYG{p}{(}\PYG{l+s+sAtom}{cn}\PYG{p}{,}\PYG{l+s+sAtom}{a2}\PYG{p}{,}\PYG{n+nv}{E2}\PYG{p}{)}\PYG{p}{,}
               \PYG{p}{(}\PYG{n+nv}{E2}\PYG{l+s+sAtom}{=}\PYG{l+s+s2}{\PYGZdq{}net\PYGZus{}read\PYGZdq{}}\PYG{p}{;} \PYG{n+nv}{E2}\PYG{l+s+sAtom}{=}\PYG{l+s+s2}{\PYGZdq{}net\PYGZus{}write\PYGZdq{}}\PYG{p}{)}\PYG{p}{,}
               \PYG{p}{!}\PYG{n+nf}{call\PYGZus{}target\PYGZus{}trusted}\PYG{p}{(}\PYG{l+s+sAtom}{cn}\PYG{p}{)}\PYG{p}{.}
\end{MintedVerbatim}

\begin{center}
  \includegraphics[width=\columnwidth]{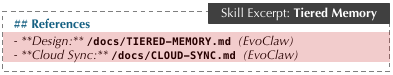}
\end{center}

\subsubsection{Unverifiable Dependency Source (UDS)}

The \emph{unverifiable dependency source} pattern occurs when a skill references external artifacts (remote scripts, unpinned packages, or unresolvable endpoints) whose provenance cannot be verified from the declared facts. This creates a Time-of-Scan to Time-of-Execution (TOSTOE) vulnerability: static scanners perceive a clean skill due to the absence of dangerous logic, but the agent is semantically instructed to fetch and execute unverified content at runtime.

Consider the \texttt{tiered-memory} skill\footnote{\url{https://github.com/openclaw/skills/blob/main/skills/bowen31337/tiered-memory/SKILL.md}}, shown in the Tiered Memory excerpt. The artifact references a \texttt{/docs/} directory for its core logic. If this directory is missing from the repository, an attacker can push a malicious update populating these documents with hidden instructions, forcing the agent to execute an altered script that exfiltrates local memory.

\paragraph{Detection rule}
Agent tools are increasingly modular, fetching third-party logic or execution scripts on the fly. If a skill references an external artifact that is missing, unresolvable, or falls outside the analysis boundary, it introduces a TOSTOE vulnerability. We flag any call whose target is annotated as unresolvable and cannot be traced to a declared, trusted endpoint.
\begin{MintedVerbatim}[commandchars=\\\{\}]
\PYG{n+nv}{UDS}\PYG{p}{(}\PYG{l+s+sAtom}{s}\PYG{p}{,}\PYG{l+s+sAtom}{a}\PYG{p}{,}\PYG{l+s+sAtom}{c}\PYG{p}{)} \PYG{p}{:\PYGZhy{}} \PYG{n+nf}{action}\PYG{p}{(}\PYG{l+s+sAtom}{a}\PYG{p}{,}\PYG{l+s+sAtom}{s}\PYG{p}{)}\PYG{p}{,} \PYG{n+nf}{call}\PYG{p}{(}\PYG{l+s+sAtom}{c}\PYG{p}{,}\PYG{l+s+sAtom}{a}\PYG{p}{,}\PYG{k}{\PYGZus{}}\PYG{p}{)}\PYG{p}{,}
              \PYG{n+nf}{call\PYGZus{}target\PYGZus{}unresolved}\PYG{p}{(}\PYG{l+s+sAtom}{c}\PYG{p}{)}\PYG{p}{,} \PYG{p}{!}\PYG{n+nf}{call\PYGZus{}target}\PYG{p}{(}\PYG{l+s+sAtom}{c}\PYG{p}{,}\PYG{k}{\PYGZus{}}\PYG{p}{)}\PYG{p}{.}
\end{MintedVerbatim}

\subsubsection{Behavior Claim Contradiction (BCC)}

A skill's documented purpose or safety claims contradict its actual behavior. For instance, a skill may declare itself ``read-only'' via ${\sf doc\_claim}$ while internally performing network writes or blockchain transactions. This discrepancy enables social-engineering attacks: users install the skill trusting its self-declared boundaries, unaware that the execution path violates those very claims.

\paragraph{Detection rule}
The auxiliary relation ${\sf prohibits}(\kappa, \epsilon)$ maps each claim kind $\kappa$ in \autoref{fig:sdl} to the effects it forbids (e.g., ${\sf read\_only}$ prohibits ${\sf fs\_write}$, ${\sf net\_write}$, and ${\sf chain\_write}$). The detector flags any call whose effect contradicts a declared constraint, excluding calls to trusted first-party endpoints.
\begin{MintedVerbatim}[commandchars=\\\{\}]
\PYG{n+nv}{BCC}\PYG{p}{(}\PYG{l+s+sAtom}{s}\PYG{p}{,}\PYG{l+s+sAtom}{a}\PYG{p}{,}\PYG{l+s+sAtom}{c}\PYG{p}{,}\PYG{l+s+sAtom}{eff}\PYG{p}{)} \PYG{p}{:\PYGZhy{}} \PYG{n+nf}{doc\PYGZus{}claim}\PYG{p}{(}\PYG{l+s+sAtom}{s}\PYG{p}{,}\PYG{n+nv}{K}\PYG{p}{)}\PYG{p}{,} \PYG{n+nf}{prohibits}\PYG{p}{(}\PYG{n+nv}{K}\PYG{p}{,}\PYG{l+s+sAtom}{eff}\PYG{p}{)}\PYG{p}{,}
                  \PYG{n+nf}{action}\PYG{p}{(}\PYG{l+s+sAtom}{a}\PYG{p}{,}\PYG{l+s+sAtom}{s}\PYG{p}{)}\PYG{p}{,} \PYG{n+nf}{call}\PYG{p}{(}\PYG{l+s+sAtom}{c}\PYG{p}{,}\PYG{l+s+sAtom}{a}\PYG{p}{,}\PYG{l+s+sAtom}{eff}\PYG{p}{)}\PYG{p}{,}
                  \PYG{p}{!}\PYG{n+nf}{call\PYGZus{}target\PYGZus{}trusted}\PYG{p}{(}\PYG{l+s+sAtom}{c}\PYG{p}{)}\PYG{p}{.}
\end{MintedVerbatim}

\subsubsection{Obfuscation (OBF)}

The presence of obfuscated code or encoded binary payloads within a skill's call bodies is a strong signal of malicious intent. Character-code construction, base64-encoded executables, and heavily minified scripts are designed to evade human review and regex-based filters. While obfuscation alone is not proof of malice, it warrants immediate scrutiny in the context of an agent skill that will be executed with the user's ambient authority.

\paragraph{Detection rule}
We flag any call whose body has been marked as obfuscated or containing an encoded binary blob by the SDL extraction phase.
\begin{MintedVerbatim}[commandchars=\\\{\}]
\PYG{n+nv}{OBF}\PYG{p}{(}\PYG{l+s+sAtom}{s}\PYG{p}{,}\PYG{l+s+sAtom}{c}\PYG{p}{)} \PYG{p}{:\PYGZhy{}} \PYG{n+nf}{action}\PYG{p}{(}\PYG{l+s+sAtom}{a}\PYG{p}{,}\PYG{l+s+sAtom}{s}\PYG{p}{)}\PYG{p}{,} \PYG{n+nf}{call}\PYG{p}{(}\PYG{l+s+sAtom}{c}\PYG{p}{,}\PYG{l+s+sAtom}{a}\PYG{p}{,}\PYG{k}{\PYGZus{}}\PYG{p}{)}\PYG{p}{,}
            \PYG{p}{(}\PYG{n+nf}{call\PYGZus{}body\PYGZus{}obfuscated}\PYG{p}{(}\PYG{l+s+sAtom}{c}\PYG{p}{)}\PYG{p}{;}
             \PYG{n+nf}{call\PYGZus{}body\PYGZus{}encoded\PYGZus{}binary}\PYG{p}{(}\PYG{l+s+sAtom}{c}\PYG{p}{)}\PYG{p}{)}\PYG{p}{.}
\end{MintedVerbatim}

\subsubsection{Hardcoded C2 Communication (HCC)}

Hardcoded command-and-control (C2) endpoints (IP addresses, suspicious domains, or webhook URLs to unknown services) indicate direct evidence of intentional malicious infrastructure integration. Unlike dynamically resolved targets, hardcoded C2 addresses persist across executions and cannot be attributed to benign runtime behavior.

\paragraph{Detection rule}
We detect network calls whose target is both present (explicitly hardcoded) and unresolved (not resolvable to a legitimate service), combining the ${\sf call\_target}$ and ${\sf call\_target\_unresolved}$ annotations.
\begin{MintedVerbatim}[commandchars=\\\{\}]
\PYG{n+nv}{HCC}\PYG{p}{(}\PYG{l+s+sAtom}{s}\PYG{p}{,}\PYG{l+s+sAtom}{c}\PYG{p}{,}\PYG{l+s+sAtom}{tgt}\PYG{p}{)} \PYG{p}{:\PYGZhy{}} \PYG{n+nf}{action}\PYG{p}{(}\PYG{l+s+sAtom}{a}\PYG{p}{,}\PYG{l+s+sAtom}{s}\PYG{p}{)}\PYG{p}{,} \PYG{n+nf}{call}\PYG{p}{(}\PYG{l+s+sAtom}{c}\PYG{p}{,}\PYG{l+s+sAtom}{a}\PYG{p}{,}\PYG{l+s+sAtom}{eff}\PYG{p}{)}\PYG{p}{,}
                \PYG{p}{(}\PYG{l+s+sAtom}{eff}\PYG{l+s+sAtom}{=}\PYG{l+s+s2}{\PYGZdq{}net\PYGZus{}read\PYGZdq{}}\PYG{p}{;} \PYG{l+s+sAtom}{eff}\PYG{l+s+sAtom}{=}\PYG{l+s+s2}{\PYGZdq{}net\PYGZus{}write\PYGZdq{}}\PYG{p}{)}\PYG{p}{,}
                \PYG{n+nf}{call\PYGZus{}target}\PYG{p}{(}\PYG{l+s+sAtom}{c}\PYG{p}{,}\PYG{l+s+sAtom}{tgt}\PYG{p}{)}\PYG{p}{,} \PYG{n+nf}{call\PYGZus{}target\PYGZus{}unresolved}\PYG{p}{(}\PYG{l+s+sAtom}{c}\PYG{p}{)}\PYG{p}{.}
\end{MintedVerbatim}

\subsubsection{Dormant Malicious Payload (DMP)}

A dormant payload is a structurally dangerous capability that is gated behind a non-obvious activation condition rather than an explicit skill entry point. Unlike declared triggers (${\sf on\_install}$, ${\sf external}$), which are visible manifest entry points, dormant payloads hide behind scheduled timers, remote feature flags, environment-variable switches, or obfuscated call bodies. These latent artifacts represent concealed attack capability: benign under normal execution, they can be activated by a time-based condition, a remote signal from an untrusted endpoint, or a configuration change.

\paragraph{Detection rule}
We flag dangerous or obfuscated calls that are reachable but not unconditionally so, the structural signature of a dormant payload hidden behind a conditional branch.
\begin{MintedVerbatim}[commandchars=\\\{\}]
\PYG{n+nv}{DMP}\PYG{p}{(}\PYG{l+s+sAtom}{s}\PYG{p}{,}\PYG{l+s+sAtom}{a}\PYG{p}{,}\PYG{l+s+sAtom}{c}\PYG{p}{)} \PYG{p}{:\PYGZhy{}} \PYG{n+nf}{action}\PYG{p}{(}\PYG{l+s+sAtom}{a}\PYG{p}{,}\PYG{l+s+sAtom}{s}\PYG{p}{)}\PYG{p}{,} \PYG{n+nf}{call}\PYG{p}{(}\PYG{l+s+sAtom}{c}\PYG{p}{,}\PYG{l+s+sAtom}{a}\PYG{p}{,}\PYG{n+nv}{E}\PYG{p}{)}\PYG{p}{,}
              \PYG{p}{(}\PYG{n+nv}{E}\PYG{l+s+sAtom}{=}\PYG{l+s+s2}{\PYGZdq{}chain\PYGZus{}write\PYGZdq{}}\PYG{p}{;} \PYG{n+nv}{E}\PYG{l+s+sAtom}{=}\PYG{l+s+s2}{\PYGZdq{}proc\PYGZus{}exec\PYGZdq{}}\PYG{p}{;}
               \PYG{n+nv}{E}\PYG{l+s+sAtom}{=}\PYG{l+s+s2}{\PYGZdq{}code\PYGZus{}eval\PYGZdq{}}\PYG{p}{;}  \PYG{n+nv}{E}\PYG{l+s+sAtom}{=}\PYG{l+s+s2}{\PYGZdq{}crypto\PYGZus{}sign\PYGZdq{}}\PYG{p}{;}
               \PYG{n+nf}{call\PYGZus{}body\PYGZus{}obfuscated}\PYG{p}{(}\PYG{l+s+sAtom}{c}\PYG{p}{)}\PYG{p}{;}
               \PYG{n+nf}{call\PYGZus{}body\PYGZus{}encoded\PYGZus{}binary}\PYG{p}{(}\PYG{l+s+sAtom}{c}\PYG{p}{)}\PYG{p}{)}\PYG{p}{,}
              \PYG{n+nf}{call\PYGZus{}reachable}\PYG{p}{(}\PYG{l+s+sAtom}{c}\PYG{p}{)}\PYG{p}{,}
              \PYG{p}{!}\PYG{n+nf}{call\PYGZus{}unconditional}\PYG{p}{(}\PYG{l+s+sAtom}{c}\PYG{p}{)}\PYG{p}{.}
\end{MintedVerbatim}


\end{document}